\documentclass[aps,floats]{revtex4}
\usepackage{amsmath,amssymb}
\usepackage{graphicx,epsfig}
\usepackage[greek,english]{babel}
\usepackage{bbold}
\makeatletter\AtBeginDocument{\let\@elt\relax}\makeatother

\DeclareMathOperator{\erf}{erf}
\DeclareMathOperator{\erfc}{erfc}

\begin{document}
\bibliographystyle {plain}

\pdfoutput=1
\def\oppropto{\mathop{\propto}} 
\def\opsimeq{\mathop{\simeq}}
\def\opoverderline{\mathop{\overline}}
\def\operarrow{\mathop{\longrightarrow}}
\def\opsim{\mathop{\sim}}

\def\fig#1#2{\includegraphics[height=#1]{#2}}
\def\figx#1#2{\includegraphics[width=#1]{#2}}


\title{ Conditioning diffusion processes with killing rates  } 


\author{Alain Mazzolo}
\affiliation{Universit\'e Paris-Saclay, CEA, Service d'\'Etudes des R\'eacteurs et de Math\'ematiques Appliqu\'ees, 91191, Gif-sur-Yvette, France}

\author{C\'ecile Monthus}
\affiliation{Universit\'e Paris-Saclay, CNRS, CEA, Institut de Physique Th\'eorique, 91191 Gif-sur-Yvette, France}


\begin{abstract}
When the unconditioned process is a diffusion submitted to a space-dependent killing rate $k(\vec x)$,
various conditioning constraints can be imposed for a finite time horizon $T$. We first analyze the conditioned process when one imposes both the surviving distribution at time $T$ and the killing-distribution for the intermediate times $t \in [0,T]$. When the conditioning constraints are less-detailed than these full distributions, we construct the appropriate conditioned processes via the optimization of the dynamical large deviations at Level 2.5 in the presence of the conditioning constraints that one wishes to impose. Finally, we describe various conditioned processes for the infinite horizon $T \to +\infty$. This general construction is then applied to two illustrative examples in order to generate stochastic trajectories satisfying various types of conditioning constraints : 
the first example concerns the pure diffusion in dimension $d$ with the quadratic killing rate $k(\vec x)= \gamma \vec x^2$, while the second example is the Brownian motion with uniform drift submitted to the delta killing rate $k(x)=k \delta(x)$ localized at the origin $x=0$.

\end{abstract}

\maketitle


\section{ Introduction }

The Doob theory of conditioned stochastic propcesses \cite{refDoob,refbookDoob}
plays an important role both in mathematics  \cite{refbookKarlin,refbookRogers,Borodin}
and in physics (see the recent review \cite{refMajumdarOrland} and references therein),
with applications in very different fields like ecology \cite{refHorne}, finance \cite{refBrody} or
nuclear engineering \cite{refMulatier,refbookPazsit}. 
For diffusion processes, many different conditioning constraints have been studied 
besides the Brownian Bridge, in particular
the Brownian excursion \cite{refMajumdarExcursion,refChung}, the Brownian meander \cite{refMajumdarMeander},
the taboo processes \cite{refKnight,refPinsky,refKorzeniowski,refGarbaczewski,refAdorisio,refAlainTaboo},
or non-intersecting Brownian bridges \cite{grela}.
Stochastic bridges have been also studied for many other Markov processes, 
including various diffusions processes \cite{henri,refSzavits,delarue},
discrete-time random walks and L\'evy flights \cite{refGarbaczewski_Levy,bruyne_discrete,Aguilar}, 
continuous-time Markov jump processes \cite{Aguilar},
run-and-tumble trajectories \cite{bruyne_run}, or
processes with resetting \cite{refdeBruyne2022}.
The stochastic bridge problem has been also extended to study the conditioning with respect to some global dynamical constraint as measured by a time-additive observable of the stochastic trajectories
\cite{refMazzoloJstat,Alain_OU,refdeBruyne2021,c_microcanonical,us_LocalTime}.

In the field of diffusion processes with killing rates that appear in many contexts \cite{Myers,Berman,Holcman,Toste}, the analysis of their conditioning has also a long history
\cite{refbookKarlin,Karlin1982,Karlin1983,Frydman,Steinsaltz,kolb,Evans2019}.
However, as discussed in detail in the recent analysis of Schr\"odinger bridges with unbalanced marginals
\cite{tryphon_killing}, the literature can be split into two groups based on different frameworks and different assumptions :
the first framework involves a Feynman-Kac multiplicative reweighting of the measure, 
while the second framework considers that one should introduce the appropriate conditioned killing rate.
As explained in detail in \cite{tryphon_killing}, the second framework is the only one compatible with 
the optimization of the relative entropy between the conditioned process and the unconditioned process
in the presence of the particular conditions that one wishes to impose.
This conclusion confirms once again the visionary perspective of E. Schr\"odinger 
in his famous paper of 1931 \cite{Schrodinger} (see the recent english translation \cite{CommentSchrodinger} with the corresponding detailed commentary), where the conditioning conditions 
are considered as the atypical result of
an experiment concerning a large number $N$ of unconditioned processes.
Via this point of view, the theory of Doob conditioning becomes connected
to the theories of dynamical large deviations and of stochastic control, as explained in detail in the commentary \cite{CommentSchrodinger},
as well as in the two reviews \cite{ControlSchrodinger,MongeSchrodinger} written from the viewpoint of stochastic control. 

The goal of the present paper is thus to revisit the conditioning of diffusion processes 
with space-dependent killing rates, in order to give a global discussion of the various conditioning constraints 
that can be imposed for finite horizon $T$ or for infinite horizon $T=+\infty$. It will be also interesting 
to see the similarities and the differences with the cases
where the diffusion process is killed only via an absorbing boundary condition
 \cite{refBaudoin,refMultiEnds,us_DoobFirstPassage,us_DoobFirstEncounter}.

The paper is organized as follows.
Section \ref{sec_unconditioned} describes the properties of the 
unconditioned diffusion processes with killing rates.
In section \ref{sec_finitehorizon}, we study the conditioning associated 
to the finite horizon $T$ where one imposes the surviving distribution $P^*(\vec y,T ) $ at time $T$
and the killing-distribution $K^*(\vec x_d,t_d ) $  for the intermediate times $t_d \in [0,T]$.
In section \ref{sec_lessdetailed}, we analyze the conditioned processes when 
the conditioning constraints are less-detailed than the full distributions $\left[ P^*(.,T) ; K^*(.,.)  \right] $ associated to the finite horizon $T$, and we discuss the consequences for the limit of the infinite horizon $T \to +\infty$.
This general formalism is then applied to the pure diffusion with quadratic killing rate in section \ref{sec_quadratic},
and to the Brownian motion with uniform drift and delta killing rate in section \ref{sec_delta},
in order to generate stochastic trajectories satisfying different conditioning constraints.
Our conclusions are summarized in section \ref{sec_conclusion}.
In Appendix \ref{app_largedev}, we describe the link with the dynamical large deviations at Level 2.5 
and the stochastic control theory. Monte Carlo simulations, also presented, illustrate our theoretical findings. 


\section{ Unconditioned process : diffusion $ \vec X(t)$ with the killing rate $k (\vec x)$ }

\label{sec_unconditioned}

In this paper, we assume that the unconditioned process $ \vec X(t)$ 
satisfies the Ito Stochastic Differential Equation (SDE) 
involving the drift $ \vec \mu (\vec x)$ and the diffusion coefficient $D (\vec x)$, 
but can also be killed with the killing rate $k (\vec x)$
\begin{eqnarray}
\vec X(t+dt)  = 
 \left\lbrace
  \begin{array}{lll}
    \emptyset  
    &~~\mathrm{with~~probability~~} k( \vec X(t)) dt
    \\
     \vec X(t) +  \vec \mu(  \vec X(t)) dt + \sqrt{ 2 D(  \vec X(t)) } d \vec W(t)
    &~~\mathrm{with~~probability~~} \left( 1-k( \vec X(t)) dt \right)
  \end{array}
\right.
\label{Itokilling}
\end{eqnarray}
where the components of $\vec W(t)$ are $d$ independent Wiener processes.
In this section, we recall the properties 
that will be useful in the other sections to construct conditioned processes.


\subsection{ Forward and backward dynamics of the propagator $P( \vec x_2,t_2 \vert  \vec x_1,t_1)$}

The generator ${\cal F} $ involving the drift $ \vec \mu (\vec x)$, the diffusion coefficient $D (\vec x)$
and the killing rate $k (\vec x)$
\begin{eqnarray}
{\cal F} \equiv  \vec \mu (\vec x) . \vec \nabla + D (\vec x)   \Delta - k (\vec x)
\label{generator}
\end{eqnarray}
governs the backward dynamics of the propagator $P( \vec x_2,t_2 \vert  \vec x,t)$
 with respect to its initial variables $( \vec x,t)$ 
 \begin{eqnarray}
-\partial_{t} P( \vec x_2,t_2 \vert  \vec x,t)  && = {\cal F} P( \vec x_2,t_2 \vert  \vec x,t)
\nonumber \\
&& =  \vec \mu( \vec x) .  \vec \nabla P( \vec x_2,t_2 \vert  \vec x,t) + D( \vec x)  \Delta P( \vec x_2,t_2 \vert  \vec x,t)
- k( \vec x) P( \vec x_2,t_2 \vert  \vec x,t) 
\label{backward}
\end{eqnarray}
while the adjoint operator of Eq. \ref{generator}
\begin{eqnarray}
{\cal F}^{\dagger} = -   \vec \nabla .  \vec \mu (\vec x)+   \Delta  D (\vec x) - k (\vec x)
\label{adjoint}
\end{eqnarray}
governs the forward dynamics of the propagator $P( \vec x,t \vert  \vec x_1,t_1)$
 with respect to its final variables $( \vec x,t)$
\begin{eqnarray}
\partial_{t} P( \vec x,t \vert  \vec x_1,t_1) 
= -   \vec \nabla . \left[   \vec \mu( \vec x) P( \vec x,t \vert  \vec x_1,t_1) \right]
+  \Delta \left[ D( \vec x) P( \vec x,t \vert  \vec x_1,t_1) \right]
- k( \vec x) P( \vec x,t \vert  \vec x_1,t_1) 
\label{forward}
\end{eqnarray}
As explained in textbooks (see for instance \cite{gardiner}), 
whenever the diffusion coefficient $D( \vec x) $ depends on the position $\vec x$,
 this forward dynamics for the propagator $P( \vec x,t \vert  \vec x_1,t_1)  $
 will be translated into two different Stochastic Differential Equations
 if one follows the Ito or the Stratonovich prescriptions for the time-discretization.
 In the present paper, we have chosen the Ito prescription
  to write the Ito-SDE of Eq. \ref{Itokilling} corresponding to Eq. \ref{forward}.
  If one wishes to use instead the Stratonovich prescription, 
  one just needs to write the corresponding Stratonovich-SDE
  that is equivalent to the forward dynamics of Eq. \ref{forward} for the propagator,
since all the forthcoming analysis will be based on 
the forward and backward dynamics of Eqs \ref{backward} and \ref{forward},
i.e. on the generator ${\cal F} $ and its adjoint $ {\cal F}^{\dagger}$.


\subsection{ Survival probability $S(T \vert  \vec x,t) $ and killing probability $K( \vec x_T, T \vert  \vec x,t )$ }

The survival probability $S(T \vert  \vec x,t) $ at time $T$ 
can be computed via the integration of the propagator $P( \vec x_T,T \vert  \vec x,t)  $ 
over the final position $ \vec x_T$ at time $T$
\begin{eqnarray}
S(T \vert  \vec x,t) = \int d^d \vec x_T P( \vec x_T,T \vert  \vec x,t) 
\label{survival}
\end{eqnarray}
Using the forward dynamics of Eq. \ref{forward},
one obtains that its time-decay
\begin{eqnarray}
- \partial_{T} S(T \vert  \vec x,t)  && = - \int d^d \vec x_T \partial_{T} P( \vec x_T,T \vert  \vec x,t) 
\nonumber \\
&& =  \int d^d \vec x_T k( \vec x_T) P( \vec x_T,T \vert  \vec x,t) 
\equiv   \int d^d \vec x_T K( \vec x_T,T \vert  \vec x,t)
\label{survivalforward}
\end{eqnarray}
involves the killing probability $K( \vec x_T,T\vert  \vec x,t ) $ at position $ \vec x_T$ at time $T$ 
\begin{eqnarray}
K( \vec x_T,T \vert  \vec x,t ) \equiv k( \vec x_T) P( \vec x_T,T \vert  \vec x,t) 
\label{killing}
\end{eqnarray}
Taking into account the initial condition $S(t \vert  \vec x,t)=1$,
the survival probability of Eq. \ref{survival} can be rewritten
\begin{eqnarray}
S(T \vert  \vec x,t) 
= 1 - \int_{t}^{T} d t_d  \int d^d \vec  x_d K(\vec x_d,t_d \vert  \vec x,t)
\label{survivalfromdead}
\end{eqnarray}
in terms of the integration of the killing probability $K(\vec x_d,t_d \vert  \vec x,t)$ 
over the position $\vec x_d$ and over the time $t_d \in [t,T]$.

The survival probability $S(T \vert  \vec x,t) $ of Eq. \ref{survival}
inherits from the propagator  $P( \vec x_T,T \vert  \vec x,t)  $
 the backward dynamics of Eq. \ref{backward} with respect to the initial variables $( \vec x,t)$
\begin{eqnarray}
-\partial_{t} S(T  \vert  \vec x,t)  = {\cal F} S(T  \vert  \vec x,t)
=  \vec \mu( \vec x) . \vec \nabla S(T   \vert  \vec x,t) + D( \vec x)  \Delta S(T  \vert  \vec x,t)
- k( \vec x) S(T  \vert  \vec x,t)
\label{Sbackward}
\end{eqnarray}
In the limit of the infinite horizon $T \to +\infty$, the forever survival probability $S(\infty \vert  \vec x)  $ 
when starting at $\vec x$ satisfies the time-independent backward equation
\begin{eqnarray}
0  = {\cal F} S(\infty \vert  \vec x) 
=  \vec \mu( \vec x) . \vec \nabla S(\infty \vert  \vec x)  + D( \vec x)  \Delta S(\infty \vert  \vec x) 
- k( \vec x) S(\infty \vert  \vec x) 
\label{Sbackwardinfty}
\end{eqnarray}


\section{ Conditioned process $\vec X^*(t)$ with respect to the finite horizon $T$ }

\label{sec_finitehorizon}

\subsection{ Conditioning towards the distribution $P^*(\vec y,T ) $ at $T$
and the killing-distribution $K^*(\vec x_d,t_d ) $  for $t_d \in [0,T]$ }

For the unconditioned diffusion process $ \vec X(t)$ starting at position $\vec X(0)=\vec x_0$ at time $t=0$ :

(i) the probability to be surviving at time $T$ at the position $\vec y$ is given by the 
unconditioned propagator $P(\vec y,T \vert \vec x_0, 0)$,
with the corresponding survival probability at time $T$ of Eq. \ref{survival}
\begin{eqnarray}
S(T \vert \vec x_0,0) = \int d^d \vec y P(\vec y,T \vert \vec x_0, 0) 
\label{survivalT}
\end{eqnarray}

 (ii) the probability to have been killed at position $\vec x_d$ at the time $t_d$ 
 is given by the unconditioned killing probability $K(\vec x_d,t_d \vert \vec x_0,0) $ of Eq. \ref{killing}
 where the corresponding probability to be already dead at time $T$
 is complementary to the survival probability of Eq. \ref{survivalT} as explained in Eq. \ref{survivalfromdead}
\begin{eqnarray}
 \int_{0}^T dt_d \int d^d \vec  x_d K(\vec x_d,t_d \vert \vec x_0,0)   = 1- S(T \vert \vec x_0,0) 
\label{deadT}
\end{eqnarray}

Now we wish to construct the conditioned diffusion process $X^*(t)$ by imposing instead 
the following other properties: 

(i) another probability $P^*(\vec y,T )$ to be surviving at position $y$ at time $T$,
whose normalization over $\vec y$ corresponds to the conditioned survival probability 
$S^*(T )$ at time $T$
\begin{eqnarray}
\int d^d \vec y P^*(\vec y,T )    = S^*(T )  
\label{survivalTstar}
\end{eqnarray}

 (ii) another probability $K^*(\vec x_d,t_d )$ 
 to have been been killed at position $\vec x_d$ at the time $t_d$, 
  whose normalization over $t_d \in [0,T]$ and over $\vec x_d$ 
  is complementary to Eq. \ref{survivalTstar}
\begin{eqnarray}
\int_0^T d t_d  \int d^d \vec  x_d K^*(\vec x_d,t_d )   = 1- S^*(T ) 
\label{deadTstar}
\end{eqnarray}
 At any intermediate time $t \in ]0,T[$, the conditioned survival probability $S^*(t)$ is given by
\begin{eqnarray}
S^*(t ) = 1-  \int_{0}^t dt_d  \int d^d \vec  x_d K^*(\vec x_d,t_d )  
\label{survivalstarinter}
\end{eqnarray}


\subsection{ Conditioned probability $P^*( \vec x,t ) $ at any intermediate time $t \in ]0,T[$ }

At any intermediate time $t \in ]0,T[$, 
the conditioned probability $P^*( \vec x,t ) $ to be surviving at position $x$
contains two contributions
\begin{eqnarray}
P^*(\vec x,t) && =  \int_t^{T} dt_d  \int d^d \vec  x_d K^*(\vec x_d,t_d )
  \frac{P(\vec x_d,t_d \vert \vec x,t) P( \vec x,t \vert \vec x_0,0)}{P(\vec x_d,t_d \vert \vec x_0,0)} 
 +  \int d^d \vec y P^*(\vec y,T )  \frac{ P(\vec y,T \vert \vec x,t) P( \vec x,t \vert \vec x_0,0)}{P(\vec y,T \vert \vec x_0,0) }
 \nonumber \\
 && \equiv Q_T(\vec x,t) P( \vec x,t \vert \vec x_0,0)
\label{conditional}
\end{eqnarray}
where $P( \vec x,t \vert \vec x_0,0) $ is the unconditioned propagator,
while the function $Q_T(\vec x,t) $ 
reads 
\begin{eqnarray}
Q_T(\vec x,t)  \equiv 
 \int_t^{T} dt_d  \int d^d \vec  x_d 
  \frac{K^*(\vec x_d,t_d ) }{P(\vec x_d,t_d \vert \vec x_0,0)} P(\vec x_d,t_d \vert \vec  x,t)
 +  \int d^d \vec y  \frac{  P^*(\vec y,T ) }{P(\vec y,T \vert \vec x_0,0) }P(\vec y,T \vert \vec  x,t)
 \label{Qdef}
\end{eqnarray}
The derivative with respect to the time $t$ appearing as the lower boundary of the integral of the first
contribution gives the following term involving the killing-time $t_d=t$
\begin{eqnarray}
- \int d^d \vec  x_d   \frac{K^*(\vec x_d,t ) }{P(\vec x_d,t \vert \vec x_0,0)} P(\vec x_d,t \vert \vec  x,t)
= - \int d^d \vec  x_d   \frac{K^*(\vec x_d,t ) }{P(\vec x_d,t \vert \vec x_0,0)} \delta^d(\vec x_d- \vec x) = - \frac{K^*( \vec x,t ) }{P( \vec x,t \vert \vec x_0,0)}
\label{Qderiborneintegrale}
\end{eqnarray}
As a consequence, this additional inhomogenous contribution appears in 
the backward dynamics that the function $Q_T(\vec x,t)  $
 inherits from 
the unconditioned propagators $P(\vec x_d,t_d \vert \vec  x,t) $ and $P(\vec y,T \vert \vec  x,t) $ 
that satisfy Eq. \ref{backward} with respect to their initial variables $(\vec x,t)$
\begin{eqnarray}
- \partial_t  Q_T(\vec x,t)   && = {\cal F}  Q_T(\vec x,t)   +  \frac{K^*( \vec x,t ) }{P( \vec x,t \vert \vec x_0,0)}
  \nonumber \\
&&  =   
   \vec \mu (\vec x) . \vec \nabla   Q_T(\vec x,t)  + D (\vec x)  \Delta Q_T(\vec x,t) - k (\vec x) Q_T(\vec x,t) 
  +  \frac{K^*( \vec x,t ) }{P( \vec x,t \vert \vec x_0,0)}
\label{Qbackward}
\end{eqnarray}


\subsection{ Forward dynamics of the conditioned process $X^*(t)$  }

Using the forward dynamics of Eq. \ref{forward} satisfied by the unconditioned propagator $P( \vec x,t \vert \vec x_0,0) $
\begin{eqnarray}
\partial_t P( \vec x,t \vert \vec x_0,0)  =  -  \vec \nabla . \left[   \vec \mu (\vec x) P( \vec x,t \vert \vec x_0,0)  \right]+   \Delta 
\left[ D (\vec x) P( \vec x,t \vert \vec x_0,0)  \right] - k (\vec x) P( \vec x,t \vert \vec x_0,0) 
\label{forwardbis}
\end{eqnarray}
and the backward dynamics of Eq. \ref{Qbackward} satisfied by the function $Q_T(\vec x,t)$,
one obtains that the time derivative of the conditioned probability of Eq. \ref{conditional}
reads
\begin{eqnarray}
  \partial_t P^*(\vec x,t)  && = P( \vec x,t \vert \vec x_0,0)  \left[   \partial_t Q_T(\vec x,t) \right]  
+ Q_T(\vec x,t) \left[  \partial_t P( \vec x,t \vert \vec x_0,0)  \right]
\nonumber \\
&& = - \vec \nabla . \left[   \vec \mu_T^*(\vec x,t) P^*(\vec x,t)  \right] +    \Delta \left[ D (\vec x) P^*(\vec x,t)   \right]
- K^*( \vec x,t ) 
\label{conditionalderi}
\end{eqnarray}
with the conditioned drift
\begin{eqnarray}
 \vec \mu_T^*(\vec x,t)  \equiv  \vec \mu (\vec x) + 2 D (\vec x)  \vec \nabla \ln Q_T(\vec x,t) 
\label{driftdoobfp}
\end{eqnarray}
while the term involving the diffusion coefficient $D (\vec x)$ is the same as in the initial dynamics.
From the point of view of the killing contribution, 
let us stress that it is the imposed conditioned killing probability $ K^*( \vec x,t ) $ that appears
directly in the conditioned dynamics of Eq. \ref{conditionalderi}.
As a consequence, the corresponding conditioned killing rate $k_T^*(\vec x,t) $ 
should be computed from the ratio
\begin{eqnarray}
k_T^*(\vec x,t) = \frac{K^*( \vec x,t ) }{P^*(\vec x,t)  } = \frac{K^*( \vec x,t ) }{ Q_T(\vec x,t) P( \vec x,t \vert \vec x_0,0)  } 
\label{killingratestar}
\end{eqnarray}
in order to generate stochastic trajectories of the conditioned process $X^*(t)$
via the following Ito SDE with killing
\begin{eqnarray}
\vec X^*(t+dt)  = 
 \left\lbrace
  \begin{array}{lll}
    \emptyset  
    &~~\mathrm{with~~probability~~} k_T^*( \vec X^*(t),t) dt
    \\
    \vec X^*(t) +  \vec \mu_T^*( \vec X^*(t),t) dt + \sqrt{ 2 D( \vec X^*(t)) } d \vec W(t)
    &~~\mathrm{with~~probability~~} \left( 1-k_T^*(\vec X^*(t),t) dt \right)
  \end{array}
\right.
\label{Itokillingstar}
\end{eqnarray}


\section{ Conditioning less detailed than the full distributions $\left[ P^*(.,T) ; K^*(.,.)  \right] $ at $T$ }

\label{sec_lessdetailed}

In section \ref{sec_finitehorizon}, 
we have constructed the conditioned process  $\vec X^*(t)$
when the conditioning constraints correspond to the full distributions $\left[ P^*(.,T) ; K^*(.,.)  \right] $ associated to the finite horizon $T$. 
In the present section, we consider instead the cases where the conditioning constraints are less detailed,
and we discuss the limit of the infinite horizon $T \to +\infty$ when appropriate.


\subsection{ Relative entropy cost of the full conditioning conditions
$\left[ P^*(.,T) ; K^*(.,.)  \right]$
imposed at the time horizon $T$}

Let us now consider a large number $N$ of independent realizations  $\vec X_n(t)$ of the unconditioned process
 labelled by $n=1,2,..,N$ starting all at the same initial condition $\vec X_n(0)=\vec x_0$ 
 at time $t=0$.
The empirical histogram ${\hat P}(\vec y,T) $ at time $T$ of the surviving position $\vec X_n(T) $ 
\begin{eqnarray}
 {\hat P}(\vec y,T) \equiv \frac{1}{N} \sum_{n=1}^N \delta^d( \vec X_n(T) - \vec y) 
 \label{empiPTN}
\end{eqnarray}
and the empirical joint histogram ${\hat K}(\vec x,t) $ of the killing events 
satisfy the global normalization 
\begin{eqnarray}
1 = \int d^d \vec y {\hat P}(\vec y,T) +\int d^d \vec x \int_0^T dt  {\hat K}(\vec x,t)
\label{globalnormaTempi}
\end{eqnarray}

The Sanov theorem 
concerning the empirical histogram of independent identically distributed variables 
(see the reviews  on large deviations \cite{oono,ellis,review_touchette} and references therein)
yields that
the joint probability to observe the empirical surviving density  ${\hat P}(\vec y,T) $  
and the empirical killing distribution ${\hat K}(\vec x,t) $ 
satisfy the large deviation form for large $N$
\begin{eqnarray}
{\cal P}^{Sanov}_T \left[ {\hat P }(.,T) ; {\hat K}(.,.)  \right]  \opsimeq_{N \to +\infty} 
\delta \left( \int d^d \vec y {\hat P}(\vec y,T) +\int d^d \vec x \int_0^T dt {\hat K}(\vec x,t) -1\right) 
e^{- N {\cal I}^{Sanov}_T \left[ {\hat P }(.,T) ; {\hat K}(.,.)  \right] }
\label{LevelSanov}
\end{eqnarray}
where the delta function imposes the normalization constraint of Eq. \ref{globalnormaTempi},
while the Sanov rate function 
\begin{eqnarray}
{\cal I}^{Sanov}_T \left[ {\hat P }(.,T) ; {\hat K}(.,.)  \right]  = 
  \int d^d \vec y   {\hat P}(\vec y,T)  \ln \left( \frac{ {\hat P}(\vec y,T)}{ P(\vec y,T \vert \vec x_0, 0)}  \right)
    +\int d^d \vec x \int_0^T dt  {\hat K}(\vec x,t)
      \ln \left(  \frac{{\hat K}(\vec x,t)  }{ k (\vec x) P( \vec x,t \vert \vec x_0,0) }  \right)      
\label{RateSanov}
\end{eqnarray} 
corresponds to the relative entropy of the empirical distributions $[{\hat P }(\vec y,T) ;  {\hat K}(\vec x,t)] $ 
with respect to their typical values $[P(\vec y,T \vert \vec x_0, 0); k (\vec x) P( \vec x,t \vert \vec x_0,0) ]$.

 Following the Schr\"odinger perspective \cite{Schrodinger}
(see the recent detailed commentary \cite{CommentSchrodinger} accompanying its english translation,
as well as in the two reviews \cite{ControlSchrodinger,MongeSchrodinger} written from the viewpoint of stochastic control), one can interpret the conditioning conditions $P^*(\vec y,T ) $ and $K^*( \vec x,t ) $ 
imposed at the finite horizon $T$ 
as the empirical results $\left[ {\hat P }(.,T) ; {\hat K}(.,.)  \right]$ 
measured in an experiment concerning
$N$ independent unconditioned processes, with the following consequences :

(i) the Sanov rate function of Eq. \ref{RateSanov} evaluated for the imposed conditions
$\left[ P^*(.,T) ; K^*(.,.)  \right]$ at the horizon $T$
\begin{eqnarray}
{\cal I}^{Sanov}_T \left[ P^*(.,T) ; K^*(.,.)  \right]
 = 
   \int d^d \vec y   P^*(\vec y,T)  \ln \left( \frac{ P^*(\vec y,T)}{ P(\vec y,T \vert \vec x_0, 0)}  \right)
    +\int d^d \vec x \int_0^T dt   K^*(\vec x,t)
      \ln \left(  \frac{K^*(\vec x,t)  }{ k (\vec x) P( \vec x,t \vert \vec x_0,0) }  \right)     
 \label{RateSanovstar}
\end{eqnarray} 
measures how rare it is for large $N$ to see the distributions $\left[ P^*(\vec y,T) ; K^*(\vec x,t)  \right]$ 
different from their typical values distributions $\left[ P(\vec y,T \vert \vec x_0,0) ; k (\vec x) P( \vec x,t \vert \vec x_0,0)  \right] $.

(ii) the Sanov rate function ${\cal I}^{Sanov}_T \left[ P^*(.,T) ; K^*(.,.)  \right] $
 of Eq. \ref{RateSanovstar} can be used
to give some precise meaning to conditioning conditions 
that are less detailed that the whole distributions $\left[ P^*(.,T) ; K^*(.,.)  \right] $  :
one needs to optimize the Sanov rate function
in the presence of the less detailed conditioning conditions that one wishes to impose
in order to construct the appropriate conditioned process.
Various simple examples are described in the following subsections.


\subsection{ Conditioning towards the surviving distribution 
$P^*(\vec y,T) $ at the horizon $T$ alone}

\label{subsec_spatial}

If one wishes to impose only the probability $P^*(\vec y,T) $ at time $T$, together
with its corresponding survival probability
\begin{eqnarray}
 S^*(T) \equiv \int d^d \vec y P^*(\vec y,T) 
  \label{survivalspace}
\end{eqnarray}
one needs to optimize the Sanov rate function ${\cal I}^{Sanov}_T \left[ P^*(.,T) ; K^*(.,.)  \right] $ of Eq. \ref{RateSanovstar}
over the killing probability $ K^*(.,.) $ normalized to 
\begin{eqnarray}
\int d^d \vec x \int_0^T dt  K^*(\vec x,t) = 1-S^*(T)
\label{normaKstar}
\end{eqnarray}
One obtains the optimal solution
\begin{eqnarray}
K^{*opt} (\vec x,t)  = \left( \frac{1- S^*(T) }{1- S(T \vert \vec x_0,0)} \right)   k (\vec x) P( \vec x,t \vert \vec x_0,0)
 \label{gammaoptfinal}
\end{eqnarray}
and its contribution to the Sanov rate function 
\begin{eqnarray}
 \int d^d \vec x \int_0^T dt  K^{*opt}(\vec x,t)         \ln \left(  \frac{K^{*opt}(\vec x,t)  }{ k (\vec x) P( \vec x,t \vert \vec x_0,0) } \right)  
  =  \left( 1- S^*(T)  \right) 
  \ln \left( \frac{1- S^*(T) }{1- S(T \vert \vec x_0,0)} \right)
 \label{lagrangianspaceopt}
\end{eqnarray}
So the relative entropy cost of the imposed probability $P^*(\vec y,T) $ at time $T$
and of its corresponding survival probability $S^*(T) $ of Eq. \ref{survivalspace} reads
\begin{eqnarray}
 {\cal I}_T^{space} \left[ P^*(.,T) ; S^*(T)  \right] 
&& =  {\cal I}^{Sanov}_T \left[ P^*(.,T) ; K^{*opt}(.,.)  \right]
 \nonumber \\
&&  =   \int d^d \vec y   P^*(\vec y,T)  \ln \left( \frac{ P^*(\vec y,T)}{ P(\vec y,T \vert \vec x_0, 0)}  \right)
  +  \left( 1- S^*(T)  \right) 
  \ln \left( \frac{1- S^*(T) }{1- S(T \vert \vec x_0,0)} \right)
 \label{ratespacealone}
\end{eqnarray}

In addition,
one should use the optimal killing probability $K^{*opt} (\vec x,t) $ of Eq. \ref{gammaoptfinal},
so the corresponding function of Eq. \ref{Qdef} becomes
\begin{eqnarray}
Q_T^{[ P^*(.,T) ; S^*(T) ] }(\vec x,t) 
 =  \left(1- S^*(T) \right) \left( \frac{1-S(T \vert  x,t) }{1- S(T \vert \vec x_0,0)} \right)
  +  \int d^d \vec y P^*(\vec y,T ) \frac{ P(\vec y,T \vert \vec  x,t)  }{P(\vec y,T \vert \vec x_0,0) }
 \label{Qspacealone}
\end{eqnarray}
while the corresponding optimal killing rate of Eq. \ref{killingratestar}
reads
\begin{eqnarray}
k_T^*(\vec x,t) = \frac{K^{*opt}( \vec x,t ) }{P^*(\vec x,t)  } 
 = \frac{ \left( \frac{1- S^*(T) }{1- S(T \vert \vec x_0,0)} \right)    }
{ Q_T^{[ P^*(.,T) ; S^*(T) ] }(\vec x,t) } k (\vec x)
\label{killingratestarspace}
\end{eqnarray}


\subsection{ Conditioning towards the killing distribution $ K^*(\vec x,t) $ for $t \in [0,T]$ alone}

\label{subsec_killing}

If one wishes to impose only the killing distribution $ K^*(\vec x,t) $ for $t \in [0,T]$ alone, together
with its normalization 
\begin{eqnarray}
\int d^d \vec x \int_0^T dt  K^*(\vec x,t) = 1-S^*(T) 
  \label{survivaltime}
\end{eqnarray}
one needs to optimize the Sanov rate function ${\cal I}^{Sanov}_T \left[ P^*(.,T) ; K^*(.,.)  \right] $
of Eq. \ref{RateSanovstar}
over the possible spatial distribution $P^*(\vec y,T)  $ normalized to $S^*(T) $.
One obtains the optimal solution
\begin{eqnarray}
P^{*opt}(\vec y,T)  = \left( \frac{ S^*(T) }{ S(T \vert \vec x_0,0)} \right)  P(\vec y,T \vert \vec x_0, 0) 
 \label{pyoptfinal}
\end{eqnarray}
and corresponding contribution to the Sanov rate function
\begin{eqnarray}  
\int d^d \vec y   P^{*opt}(\vec y,T)  \ln \left( \frac{ P^{*opt}(\vec y,T)}{ P(\vec y,T \vert \vec x_0, 0)} \right)
  =  S^*(T) 
   \ln \left( \frac{ S^*(T) }{ S(T \vert \vec x_0,0)} \right)
 \label{lagrangiantimeopt}
\end{eqnarray}
So the relative entropy cost of the killing distribution $K^*(\vec x,t) $ 
and of the corresponding survival probability $S^*(T) $ of Eq. \ref{survivaltime}
is given by
\begin{eqnarray}
 {\cal I}_T^{killing} \left[  K^*(.,.)  ; S^*(T) \right]
 && =  {\cal I}^{Sanov}_T \left[ P^{*opt}(.,T) ; K^*(.,.)  \right]
 \nonumber \\
&&  =  \int d^d \vec x \int_0^T dt  K^*(\vec x,t)         \ln \left(  \frac{K^*(\vec x,t)  }{ k (\vec x) P( \vec x,t \vert \vec x_0,0) } \right)    
  +  S^*(T)    \ln \left( \frac{ S^*(T) }{ S(T \vert \vec x_0,0)} \right)
   \label{ratekillingalone}
\end{eqnarray}

In addition,
one should use the optimal solution $P^{*opt}(\vec y,T) $ of Eq. \ref{pyoptfinal},
so the function of Eq. \ref{Qdef} becomes
\begin{eqnarray}
Q_T^{[  K^*(.,.)  ; S^*(T) ]}(\vec x,t)  = 
 \int_t^{T} dt_d  \int d^d \vec  x_d K^*(\vec x_d,t_d )
  \frac{P(\vec x_d,t_d \vert \vec  x,t) }{P(\vec x_d,t_d \vert \vec x_0,0)} 
 +S^*(T)  \left( \frac{S(T \vert \vec  x,t) }{ S(T \vert \vec x_0,0)} \right) 
 \label{Qkillingalone}
\end{eqnarray}
with the corresponding optimal killing rate of Eq. \ref{killingratestar}
\begin{eqnarray}
k_T^*(\vec x,t) = \frac{K^*( \vec x,t ) }{Q_T^{[  K^*(.,.)  ; S^*(T) ]}(\vec x,t)   P( \vec x,t \vert \vec x_0,0)  }
\label{killingratestarkilling}
\end{eqnarray}


\subsubsection*{ Application to the conditioning towards the normalized killing distribution $ K^*(\vec x,t) $ for 
the infinite horizon $T=+\infty$ }

Let us now consider the limit of the infinite horizon $T \to +\infty$, where one wishes to impose 
some normalized killing distribution $ K^*(\vec x,t) $ in Eq. \ref{survivaltime}
\begin{eqnarray}
\int d^d \vec x \int_0^{+\infty} dt  K^*(\vec x,t) = 1-S^*(\infty) =1
  \label{survivaltimeinfty}
\end{eqnarray}
so that the conditioned forever-survival probability vanishes $S^*(\infty)=0 $.
Then the function of Eq. \ref{Qkillingalone} reduces to
\begin{eqnarray}
Q_{\infty}^{[  K^*(.,.)  ; S^*(\infty)=0 ]}(\vec x,t)  = 
 \int_t^{+\infty} dt_d  \int d^d \vec  x_d K^*(\vec x_d,t_d )
  \frac{P(\vec x_d,t_d \vert \vec  x,t) }{P(\vec x_d,t_d \vert \vec x_0,0)} 
 \label{Qkillingaloneinfty}
\end{eqnarray}
while the conditioned killing rate of Eq. \ref{killingratestarkilling} reads
\begin{eqnarray}
k_{\infty}^*(\vec x,t) = \frac{K^*( \vec x,t ) }{Q_{\infty}^{[  K^*(.,.)  ; S^*(\infty)=0 ]}(\vec x,t)   P( \vec x,t \vert \vec x_0,0)  }
\label{killingratestarkillinginfty}
\end{eqnarray}


\subsection{ Conditioning towards the surviving probability $ S^*(T) $ at time $T$ alone}

If one wishes to impose only the value $S^*(T)$ of the conditioned survival probability at time $T$,
the computations of the two previous subsections \ref{subsec_spatial} and \ref{subsec_killing}
can be used to obtain the following results.
The relative entropy cost of imposing 
the surviving probability $ S^*(T) $ at time $T$ alone reduces to
\begin{eqnarray}
 {\cal I}^{surviving}_T \left[  S^*(T) \right]
 && =        S^*(T)    \ln \left( \frac{ S^*(T) }{ S(T \vert \vec x_0,0)} \right)
   +  \left( 1- S^*(T)  \right)   \ln \left( \frac{1- S^*(T) }{1- S(T \vert \vec x_0,0)} \right)
   \label{ratesurvivalalone}
\end{eqnarray}
In addition, the optimal solutions $K^{*opt} (\vec x,t) $ of Eq. \ref{gammaoptfinal}
and $P^{*opt}(\vec y,T) $ of Eq. \ref{pyoptfinal} yield that
 the appropriate function $Q_T(\vec x,t)$ reads using Eqs \ref{Qspacealone} and \ref{Qkillingalone}
\begin{eqnarray}
Q_T^{[  S^*(T)]}(\vec x,t) 
 =  \left( \frac{1- S^*(T) }{1- S(T \vert \vec x_0,0)} \right) \left( 1-S(T \vert  x,t) \right) 
+ \left( \frac{ S^*(T) }{ S(T \vert \vec x_0,0)} \right) S(T \vert \vec  x,t)
 \label{Qsurvivingalone}
\end{eqnarray}
with the corresponding optimal killing rate of Eq. \ref{killingratestar}
\begin{eqnarray}
k_T^*(\vec x,t) = \frac{ \left( \frac{1- S^*(T) }{1- S(T \vert \vec x_0,0)} \right)    }
{ Q_T^{[  S^*(T) ] }(\vec x,t) } k (\vec x)
\label{killingratestarsurvival}
\end{eqnarray}


\subsubsection{ Special case : conditioning towards the survival probability $S^*(T)=1 $ 
at the finite horizon $T$ }

For the special case where one wishes to impose the full survival $S^*(T)=1 $ 
at the finite horizon $T$, 
the conditioned killing rate of Eq. \ref{killingratestarsurvival} vanishes as it should
\begin{eqnarray}
k_T^*(\vec x,t) = 0
\label{killingratestarsurvivalstars1}
\end{eqnarray}
while the function of Eq. \ref{Qsurvivingalone} reduces to the ratio of survival probabilities 
$S(T \vert   .,.)  $  of the unconditioned process
\begin{eqnarray}
Q_T^{[  S^*(T)=1]}(\vec x,t) 
 =  \frac{ S(T \vert \vec  x,t) }{ S(T \vert \vec x_0,0)} 
 \label{Qsurvivingalonestars1}
\end{eqnarray}
So the corresponding conditioned drift of Eq. \ref{driftdoobfp} reads
in terms of the unconditioned survival probability $S(T \vert \vec  x,t) $
\begin{eqnarray}
 \vec \mu_T^*(\vec x,t)  =  \vec \mu (\vec x) + 2 D (\vec x)  \vec \nabla \ln  S(T \vert \vec  x,t) 
\label{driftdoobfpstars1}
\end{eqnarray}
in agreement with the general formula given in \cite{Frydman}, where many illustrative examples can be found,
including an application in section 4 to an unconditioned diffusion process
on the half-line $[0,+\infty[$, where 
the drift $\mu(x)$, the diffusion coefficient $D(x)$ and the killing rate $k(x)$
are all linear functions of the position $x$.


\subsubsection{ Special case : conditioning towards the survival probability $S^*(T)=0 $ 
at the finite horizon $T$ }

For the special case where one wishes to impose no survival $S^*(T)=0 $ 
at the finite horizon $T$, 
 the function of Eq. \ref{Qsurvivingalone} reduces to
\begin{eqnarray}
Q_T^{[  S^*(T)=0]}(\vec x,t) 
 =  \frac{ 1-S(T \vert \vec  x,t) }{ 1-S(T \vert \vec x_0,0)} 
 \label{Qsurvivingalonestars0}
\end{eqnarray}
So the corresponding conditioned drift of Eq. \ref{driftdoobfp} is given by
\begin{eqnarray}
 \vec \mu_T^*(\vec x,t)  =  \vec \mu (\vec x) + 2 D (\vec x)  \vec \nabla \ln  \left[ 1-S(T \vert \vec  x,t) \right]
\label{driftdoobfpstars0}
\end{eqnarray}
while
the conditioned killing rate of Eq. \ref{killingratestarsurvival} reads
\begin{eqnarray}
k_T^*(\vec x,t)  = \frac{ k (\vec x)   }
{ \left[ 1- S(T \vert \vec x_0,0) \right] Q_T^{[  S^*(T)=0 ] }(\vec x,t) } 
= \frac{ k (\vec x)   } { 1-S(T \vert \vec  x,t)} 
\label{killingratestarsurvivalstar0}
\end{eqnarray}


\subsection{ Limit of the infinite horizon $T=+\infty$ : conditioning towards the forever-survival probability $S^*(\infty) $  }

Let us now consider the limit of the infinite horizon $T \to +\infty$, where one wishes to impose only
the forever-survival probability $S^*(\infty) $.
Then the function of Eq. \ref{Qsurvivingalone} should be computed from the limits 
\begin{eqnarray}
Q_{\infty}^{[  S^*(\infty)]}(\vec x,t) 
 =  (1-S^*(\infty)) \lim_{T \to +\infty} \left( \frac{1-S(T \vert  \vec x,t) }{1- S(T \vert \vec x_0,0)} \right)
 + S^*(\infty) \lim_{T \to +\infty} \left( \frac{S(T \vert \vec  x,t) }{ S(T \vert \vec x_0,0)} \right)
 \label{Qsurvivingaloneinfty}
\end{eqnarray}
while the conditioned killing rate of Eq. \ref{killingratestarsurvival} becomes
\begin{eqnarray}
k_{\infty}^*(\vec x,t) = \frac{ \left( \frac{1- S^*(\infty) }{1- S( \infty \vert \vec x_0)} \right)    }
{ Q_{\infty}^{[  S^*(\infty) ] }(\vec x,t) } k (\vec x)
\label{killingratestarsurvivalinfty}
\end{eqnarray}

In order to evaluate the limits of Eq. \ref{Qsurvivingaloneinfty}
one needs to distinguish whether the unconditioned
survival probability $S(T \vert .,.)$ vanishes or not for $T \to +\infty$.


\subsubsection{ Cases where the unconditioned
survival probability vanishes $S(+\infty \vert .) =0$   }

When the unconditioned survival probability vanishes $S(+\infty \vert .) =0$,
the function of Eq. \ref{Qsurvivingaloneinfty}
\begin{eqnarray}
Q_{\infty}^{[  S^*(\infty)]}(\vec x,t) 
 =  (1-S^*(\infty))
 + S^*(\infty) \lim_{T \to +\infty} \left( \frac{S(T \vert  \vec x,t) }{ S(T \vert \vec x_0,0)} \right)
 \label{Qsurvivingaloneinftyzero}
\end{eqnarray}
involves the limit of the ratio $\left( \frac{S(T \vert \vec x,t) }{ S(T \vert \vec x_0,0)} \right) $ of the two vanishing survival probabilities.


\subsubsection{  Cases where the unconditioned
survival probability remains finite $S(+\infty \vert .) \in ]0,1[$   }

When the unconditioned survival probability remains finite $S(+\infty \vert .) \in ]0,1[$,
the function of Eq. \ref{Qsurvivingaloneinfty} reduces to
\begin{eqnarray}
Q_{\infty}^{[  S^*(\infty)]}(\vec x) 
 =  (1-S^*(\infty))  \left( \frac{1-S(\infty \vert  \vec x) }{1- S(\infty \vert \vec x_0)} \right)
 + S^*(\infty)  \left( \frac{S(\infty \vert  \vec x) }{ S(\infty \vert \vec x_0)} \right)
 \label{Qsurvivingaloneinftyfinite}
\end{eqnarray}
while the conditioned killing rate of Eq. \ref{killingratestarsurvivalinfty} reads
\begin{eqnarray}
k_{\infty}^*(\vec x) = \frac{ \left( \frac{1- S^*(\infty) }{1- S( \infty \vert \vec x_0)} \right)    }
{ Q_{\infty}^{[  S^*(\infty) ] }(\vec x) } k (\vec x)
\label{killingratestarsurvivalinftyfinite}
\end{eqnarray}

An explicit example where the conditioning is towards zero-survial $S^*(\infty)=0 $
can be found in the book  \cite{refbookKarlin} on pages 282-283
for the unconditioned diffusion process on the half-line $[0,+\infty[$ with no drift $\mu(x)=0$,
where the diffusion coefficient $D(x)$ and the killing rate $k(x)$ are given by the space-dependent functions
\begin{eqnarray}
D(x) && =\frac{x}{2}
\nonumber \\
k(x) && = \frac{x^2}{2}
\label{exempleKarlin}
\end{eqnarray}


\subsection{ Conditioning towards the time-killing distribution $ K^*(t) = \int_{-\infty}^{+\infty} d^d \vec x K^*(\vec x,t) $ for $t \in ]0,T[$ alone}

If one wishes to impose only the time-killing distribution $ K^*(t) = \int d^d \vec x K^*(\vec x,t) $ for $t \in ]0,T[$, together with its normalization from Eq. \ref{survivaltime}
\begin{eqnarray}
 \int_0^T dt  K^*(t) = 1-S^*(T) 
  \label{survivaltimekilling}
\end{eqnarray}
one needs to optimize the rate function ${\cal I}_T^{killing} \left[  K^*(.,.)  ; S^*(T) \right] $
of Eq. \ref{ratekillingalone}
over the possible spatial-dependence in $x$ of the killing distributions  $K^*(\vec x,t)  $,
with the normalization constraint for each $t$
\begin{eqnarray}
\int_{-\infty}^{+\infty} d^d \vec x K^*(\vec x,t) = K^*(t)
  \label{normakillinganty}
\end{eqnarray}
One obtains the optimal solution
\begin{eqnarray}
K^{*opt}(\vec x,t)  = \frac{K^*(t)}{K(t \vert \vec x_0,0) } k (\vec x) P( \vec x,t \vert \vec x_0,0)
 \label{kstarxoptfinal}
\end{eqnarray}
where
\begin{eqnarray}
K(t \vert \vec x_0,0)  \equiv \int d^d \vec x k (\vec x) P( \vec x,t \vert \vec x_0,0)
 \label{kttyp}
\end{eqnarray}
represents the unconditioned time-killing probability.
The corresponding contribution to the rate function of Eq. \ref{ratekillingalone}
\begin{eqnarray}
 \int d^d \vec x \int_0^T dt  K^*(\vec x,t)         \ln \left(  \frac{K^*(\vec x,t)  }{ k (\vec x) P( \vec x,t \vert \vec x_0,0) } \right)    
  = \int_0^T dt K^*(t) \ln \left(  \frac{K^*(t)  }{ K(t\vert \vec x_0,0) } \right)  
\label{lagrangiantimeoptcontri}
\end{eqnarray}
leads to the following relative entropy cost 
\begin{eqnarray}
 {\cal I}_T^{time} \left[  K^*(.)  ; S^*(T) \right]
 && =  \int_0^T dt K^*(t) \ln \left(  \frac{K^*(t)  }{ K(t\vert \vec x_0,0) } \right)  
  +  S^*(T)    \ln \left( \frac{ S^*(T) }{ S(T \vert \vec x_0,0)} \right)
   \label{ratetimekillingalone}
\end{eqnarray}

In addition,
one should use the optimal solution $K^{*opt} (\vec x,t) $ of Eq. \ref{kstarxoptfinal},
so the function of Eq. \ref{Qkillingalone} becomes
\begin{eqnarray}
Q_T^{[K^*(.)  ; S^*(T) ]}(\vec x,t) && =
 \int_t^{T} dt_d   \frac{K^*(t_d)}{K(t_d\vert \vec x_0,0) } \int d^d \vec  x_d    k(\vec x_d) P(\vec x_d,t_d \vert \vec x,t)
 + \left( \frac{ S^*(T) }{ S(T \vert \vec x_0,0)} \right) S(T \vert \vec x,t) 
  \nonumber \\
&& =
 \int_t^{T} dt_d   \frac{K^*(t_d)}{K(t_d \vert \vec x_0,0) } K(t_d \vert \vec x,t)
 + \left( \frac{ S^*(T) }{ S(T \vert \vec x_0,0)} \right) S(T \vert \vec x,t) 
 \label{Qtimekillingalone}
\end{eqnarray}
with the corresponding optimal killing rate of Eq. \ref{killingratestarkilling} 
\begin{eqnarray}
k_T^*(\vec x,t) = \frac{K^{*opt}( \vec x,t ) }{Q_T^{[  K^*(.)  ; S^*(T) ]}(\vec x,t)   P( \vec x,t \vert \vec x_0,0)  }
=  \frac{ \frac{K^*(t)}{K(t \vert \vec x_0,0) } }
{Q_T^{[  K^*(.)  ; S^*(T) ]}(\vec x,t)   }  k (\vec x)
\label{killingratestarkillingtime}
\end{eqnarray}


\subsubsection*{ Application to the conditioning towards the normalized time-killing distribution $ K^*(t) $ for 
the infinite horizon $T=+\infty$ }

Let us now consider the limit of the infinite horizon $T \to +\infty$, where one wishes to impose 
some normalized time-killing distribution $ K^*(t) $ in Eq. \ref{survivaltime}
\begin{eqnarray}
 \int_0^{+\infty} dt  K^*(t) = 1-S^*(\infty) =1
  \label{survivaltimeKinfty}
\end{eqnarray}
so that the conditioned forever-survival probability vanishes $S^*(\infty)=0 $.
Then the function of Eq. \ref{Qtimekillingalone} becomes
\begin{eqnarray}
Q_{\infty}^{[K^*(.)  ; S^*(\infty) =0]}(\vec x,t) && =
 \int_t^{+\infty}  dt_d   \frac{K^*(t_d)}{K(t_d \vert \vec x_0,0) } K(t_d \vert \vec x,t)
 \label{Qtimekillingaloneinfty}
\end{eqnarray}
while the conditioned killing rate of Eq. \ref{killingratestarkillingtime} reads
\begin{eqnarray}
k_{\infty}^*(\vec x,t) 
=  \frac{ \frac{K^*(t)}{K(t \vert \vec x_0,0) } }
{Q_{\infty}^{[K^*(.)  ; S^*(\infty) =0]}(\vec x,t)    }  k (\vec x)
\label{killingratestarkillingtimeinfty}
\end{eqnarray}


\section{Application to pure diffusion with quadratic killing rate }

\label{sec_quadratic}

In this section, the general framework described previously
is applied to the explicit case where the unconditioned process is the pure diffusion in dimension $d$
with the quadratic killing rate $k (\vec x)=\gamma \vec x^2$.

\subsection{ Unconditioned process $ \vec X(t)$ : diffusion coefficient $D$ and quadratic killing rate $k (\vec x)=\gamma \vec x^2$ }

The unconditioned process $ \vec X(t)$
 is generated by Eq. \ref{Itokilling}
with no drift $ \vec \mu (\vec x)= \vec 0$, with the uniform diffusion coefficient $D (\vec x)=D$, 
and the quadratic killing rate $k (\vec x)=\gamma \vec x^2$
\begin{eqnarray}
\vec X(t+dt)  = 
 \left\lbrace
  \begin{array}{lll}
    \emptyset  
    &~~\mathrm{with~~probability~~} \gamma  \vec X^2(t)  dt
    \\
     \vec X(t)   + \sqrt{2D}  d \vec W(t)
    &~~\mathrm{with~~probability~~} \left( 1- \gamma  \vec X^2(t) dt \right)
  \end{array}
\right.
\label{Itokillingbrown}
\end{eqnarray}
The self-adjoint generator of Eq. \ref{generator}
\begin{eqnarray}
{\cal F} = {\cal F}^{\dagger} =  D    \Delta - \gamma \vec x^2 \equiv  - H
\label{generatoroh}
\end{eqnarray}
corresponds to the quantum Hamiltonian
\begin{eqnarray}
H \equiv - \frac{1}{2m} \Delta + \frac{ m \omega^2}{2} \vec x^2 
\label{hamiltonianoh}
\end{eqnarray}
of an harmonic oscillator
of mass $m=\frac{1}{2D}$ and of frequency
\begin{eqnarray}
\omega \equiv 2 \sqrt{D \gamma}
\label{omega}
\end{eqnarray}
As a consequence, the unconditioned propagator  $P( \vec x_2,t_2 \vert  \vec x_1,t_1) $
corresponds to the Euclidean propagator of the quantum harmonic oscillator
that reads
\begin{eqnarray}
P( \vec x_2,t_2 \vert  \vec x_1,t_1)  = 
\left( \frac{\omega}{4 D \pi \sinh [\omega (t_2-t_1)]} \right)^{\frac{d}{2}}  
e^{ \displaystyle -\frac{\omega}{4 D \sinh [\omega (t_2-t_1)]}
 [ (\vec x_2^2 + \vec x_1^2)\cosh [\omega (t_2-t_1)]  -2 \vec x_2. \vec x_1]} 
\label{oh}
\end{eqnarray}
The unconditioned survival probability of Eq. \ref{survival} reads
\begin{eqnarray}
S(t_2 \vert  \vec x_1,t_1) =\int d^d \vec  x_2 P( \vec x_2,t_2 \vert  \vec x_1,t_1)   
=  \frac{1}{ \cosh^{\frac{d}{2}} [\omega (t_2-t_1)]}  
e^{ \displaystyle -  \frac{\omega \tanh [\omega (t_2-t_1)]}{4 D }  \vec x_1^2}
\label{survivaloh}
\end{eqnarray}
Its decay with respect to $t_2$ gives the unconditioned killing-time probability
\begin{eqnarray}
K(t_2 \vert  \vec x_1,t_1) && = - \partial_{t_2} S(t_2 \vert  \vec x_1,t_1) 
\nonumber \\
&& =  \frac{1}{ \cosh^{\frac{d}{2}+1} [\omega (t_2-t_1)]}  
e^{ \displaystyle -  \frac{\omega \tanh [\omega (t_2-t_1)]}{4 D }  \vec x_1^2}
\left[ \frac{d}{2} \omega \sinh [\omega (t_2-t_1)] + \frac{\omega^2 }{4D \cosh [\omega (t_2-t_1)} \vec x_1^2\right]
\label{timekillingoh}
\end{eqnarray}
while the unconditioned space-time killing probability of Eq. \ref{killing} reads using Eq. \ref{omega} 
to replace $\gamma= \frac{\omega^2}{4D}$
\begin{eqnarray}
K( \vec x_2,t_2 \vert  \vec x_1,t_1 ) && =  k(\vec x_2) P( \vec x_2,t_2 \vert  \vec x_1,t_1)
\nonumber \\
&& = \frac{\omega^2}{4D} \vec x_2^2
\left( \frac{\omega}{4 D \pi \sinh [\omega (t_2-t_1)]} \right)^{\frac{d}{2}}  
e^{ \displaystyle -\frac{\omega}{4 D \sinh [\omega (t_2-t_1)]}
 [ (\vec x_2^2 + \vec x_1^2)\cosh [\omega (t_2-t_1)]  -2 \vec x_2. \vec x_1]} 
\label{killingoh}
\end{eqnarray}


\subsection{ Full conditioning constraints $\left[ P^*(.,T) ; K^*(.,.)  \right]  $ associated to the finite horizon $T$ }

Using the propagator of Eq. \ref{oh}, one obtains that the function $Q_T(\vec x,t) $ of Eq. \ref{Qdef} reads
\begin{small}
\begin{eqnarray}
&& Q_T(\vec x,t)   = 
 \int_t^{T} dt_d  \int d^d \vec  x_d K^*(\vec x_d,t_d )
  \frac{P(\vec x_d,t_d \vert \vec x,t) }{P(\vec x_d,t_d \vert \vec x_0,0)} 
 +  \int d^d \vec y P^*(\vec y,T ) \frac{ P(\vec y,T \vert \vec x,t)  }{P(\vec y,T \vert \vec x_0,0) }
 \nonumber \\
 && =  \int_t^{T} dt_d  \int d^d \vec  x_d K^*(\vec x_d,t_d )
 \left( \frac{\sinh [\omega t_d]}{ \sinh [\omega (t_d-t)]} \right)^{\frac{d}{2}}
  e^{ 
   \displaystyle 
  \frac{\omega (\vec x_d^2 + \vec x_0^2) }{4 D \tanh [\omega t_d]}
  -\frac{\omega   (\vec x_d^2 + \vec x^2)  }{4 D \tanh [\omega (t_d-t)]}   
  +\frac{\omega  \vec x_d. \vec x}{2 D \sinh [\omega (t_d-t)]}
  - \frac{\omega  \vec x_d. \vec x_0}{2 D \sinh [\omega t_d]}
 }  
 \nonumber \\
 && +  \int d^d \vec y P^*(\vec y,T ) 
 \left( \frac{\sinh [\omega T]}{ \sinh [\omega (T-t)]} \right)^{\frac{d}{2}}
  e^{ 
   \displaystyle 
 \frac{\omega (\vec y^2 + \vec x_0^2) }{4 D \tanh [\omega T]}
  -\frac{\omega   (\vec y^2 + \vec x^2)  }{4 D \tanh [\omega (T-t)]}   
  +\frac{\omega  \vec y. \vec x}{2 D \sinh [\omega (T-t)]} 
  -\frac{\omega  \vec y. \vec x_0}{2 D \sinh [\omega T]}  
 }  
 \label{Qoh}
\end{eqnarray}
\end{small}


\subsubsection*{ Example of the bridge without being killed : $P^*(\vec y,T )=\delta^d(\vec y-\vec y_*)   $  }

For the case where one imposes the full survival at time $T$ at the single position $\vec y_* $ 
\begin{eqnarray}
P^*(\vec y,T ) && =\delta^d(\vec y-\vec y_*)
\nonumber \\
K^*(\vec x_d,t_d ) && =0
 \label{pspatial1delta}
\end{eqnarray}
 Eq. \ref{Qoh} reduces to
\begin{eqnarray}
Q_T(\vec x,t) =
 \left( \frac{\sinh [\omega T]}{ \sinh [\omega (T-t)]} \right)^{\frac{d}{2}}
  e^{ 
   \displaystyle 
 \frac{\omega (\vec y_*^2 + \vec x_0^2) }{4 D \tanh [\omega T]}
  -\frac{\omega   (\vec y_*^2 + \vec x^2)  }{4 D \tanh [\omega (T-t)]}   
  +\frac{\omega  \vec y_*. \vec x}{2 D \sinh [\omega (T-t)]} 
  -\frac{\omega  \vec y_*. \vec x_0}{2 D \sinh [\omega T]}  
 }  
 \label{QohSurvivaly1delta}
\end{eqnarray}
The corresponding conditioned drift of Eq. \ref{driftdoobfp} reads
\begin{eqnarray}
\vec \mu_T^*(\vec x,t)   = 2 D   \vec \nabla \ln Q_T(\vec x,t)
=     \frac{\omega  }{  \sinh [\omega (T-t)]}  \vec y_* 
- \frac{\omega  }{  \tanh [\omega (T-t)]}   \vec x 
    \label{driftohSurvival1delta}
\end{eqnarray}
while the conditioned killing rate of Eq. \ref{killingratestar} vanishes $k_T^*(x,t)=0$.
The Brownian bridge is recovered in the limit $\omega \to 0$ as it should.

\begin{figure}[h]
\centering
\includegraphics[width=4.2in,height=3.2in]{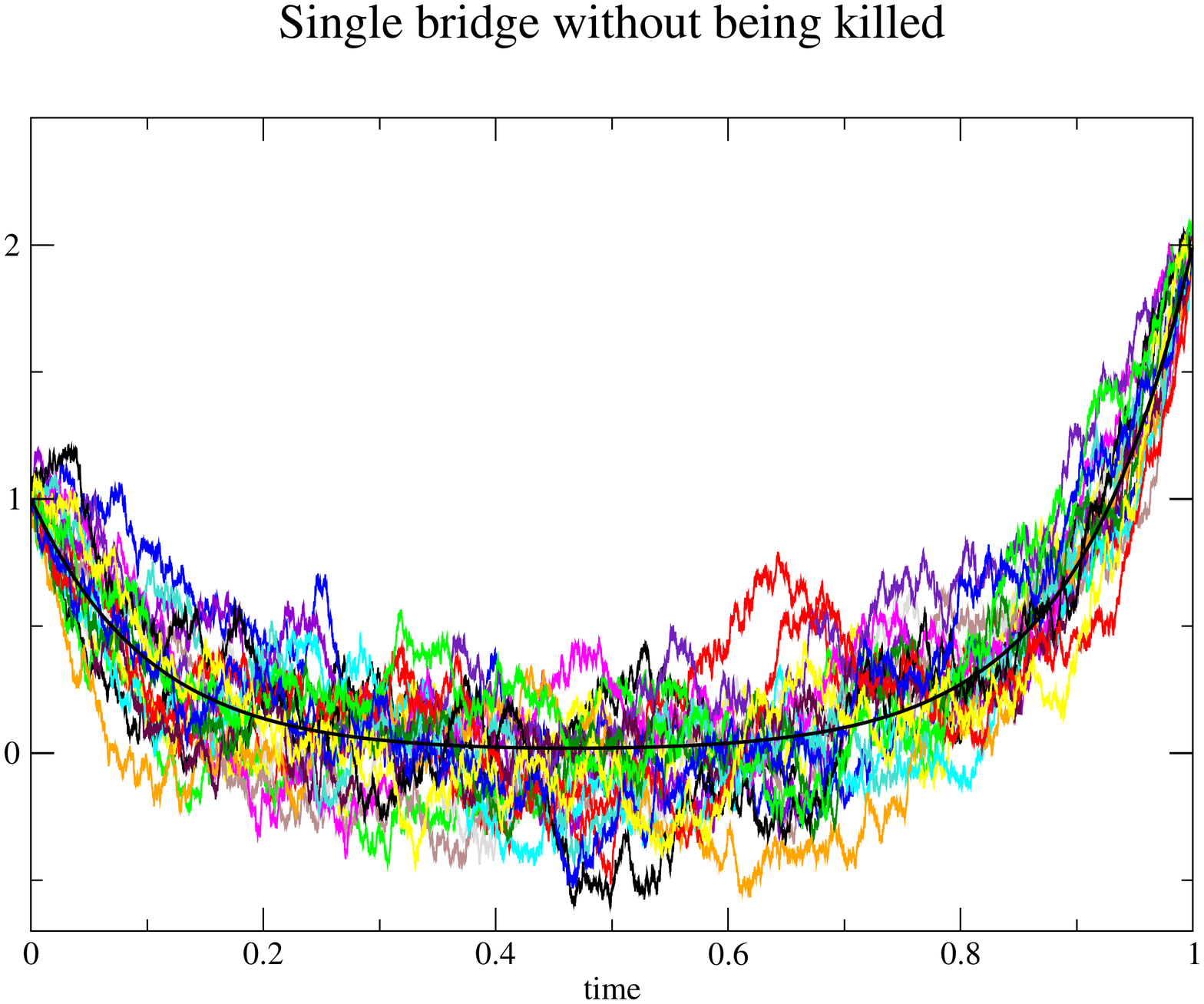}
\setlength{\abovecaptionskip}{15pt}  
\caption{A sample of diffusions satisfying the Ito stochastic differential equation Eq. \ref{ItodriftohSurvival1delta}. Each color corresponds to the realization of one  process. The thick black curve is the average profile of the stochastic process as given by equation \ref{averageItodriftohSurvival1delta}. The time step used in the discretization is $dt = 10^{-4}$. Due to the killing rate $\frac{\omega^2}{4 D}x^2$, trajectories that are likely to survive spend most of their time near $x=0$.}
\label{fig1}
\end{figure}

In dimension $d=1$, the conditioned process thus satisfies the Ito stochastic differential equation 
\begin{eqnarray}
 dX^*(t) =  \left( \frac{\omega  }{  \sinh [\omega (T-t)]}  y_* 
- \frac{\omega  }{  \tanh [\omega (T-t)]} X^*(t) \right) dt + dW(t)
 \label{ItodriftohSurvival1delta}
\end{eqnarray}
Figure \ref{fig1} shows a set of $20$ realizations of the process with parameter $\omega = 10$ as well as the mean trajectory $\langle X^*(t) \rangle$. This last quantity is obtained by averaging the preceding equation over the realizations. Since $\langle dW(t) \rangle = 0$, we have
\begin{eqnarray}
 \frac{ d \langle X^*(t)\rangle}{dt} =   \frac{\omega  }{  \sinh [\omega (T-t)]}  y_* 
- \frac{\omega  }{  \tanh [\omega (T-t)]} \langle X^*(t)\rangle 
 \label{averageeqdiffItodriftohSurvival1delta}
\end{eqnarray}
Solving the linear Eq. \ref{averageeqdiffItodriftohSurvival1delta} is straightforward and we get 
\begin{eqnarray}
 \langle X^*(t) \rangle = y^* \frac{\sinh(\omega t)}{\sinh(\omega T)} + x_0 \frac{\sinh(\omega (T-t))}{\sinh(\omega T)}
 \label{averageItodriftohSurvival1delta}
\end{eqnarray}



\subsection{ Conditioning towards the surviving distribution 
$P^*(\vec y,T) $ at the horizon $T$ alone}

If one wishes to impose only the probability $P^*(\vec y,T) $ at time $T$, together
with its corresponding survival probability $S^*(T) = \int d^d \vec y P^*(\vec y,T) $,
the function of Eq. \ref{Qspacealone} reads
\begin{eqnarray}
&& Q_T^{[ P^*(.,T) ; S^*(T) ] }(\vec x,t) = \left[1- S^*(T) \right] \left( \frac{1-S(T \vert  x,t) }
{1- S(T \vert \vec x_0,0)} \right)
  +  \int d^d \vec y P^*(\vec y,T ) \frac{ P(\vec y,T \vert \vec  x,t)  }{P(\vec y,T \vert \vec x_0,0) }
\nonumber \\
&& =  \left[1- S^*(T) \right]
\left( \frac{1-  \frac{1}{ \cosh^{\frac{d}{2}} [\omega (T-t)]}  e^{  -  \frac{\omega \tanh [\omega (T-t)]}{4 D }  \vec x^2} }
{1- \frac{1}{ \cosh^{\frac{d}{2}} [\omega T]} e^{  -  \frac{\omega \tanh [\omega T]}{4 D }  \vec x_0^2} } \right)
 \\
&&  +  \int d^d \vec y P^*(\vec y,T ) 
 \left( \frac{\sinh [\omega T]}{ \sinh [\omega (T-t)]} \right)^{\frac{d}{2}}
  e^{ 
   \displaystyle 
 \frac{\omega (\vec y^2 + \vec x_0^2) }{4 D \tanh [\omega T]}
  -\frac{\omega   (\vec y^2 + \vec x^2)  }{4 D \tanh [\omega (T-t)]}   
  +\frac{\omega  \vec y. \vec x}{2 D \sinh [\omega (T-t)]} 
  -\frac{\omega  \vec y. \vec x_0}{2 D \sinh [\omega T]}  
 }  \nonumber
 \label{Qspacealoneoh}
\end{eqnarray}


\subsection{ Conditioning towards the killing distribution $ K^*(\vec x,t) $ for $t \in [0,T]$ alone}

If one wishes to impose only the killing distribution $ K^*(\vec x,t) $ for $t \in [0,T]$ alone, together
with its normalization $ \int d^d \vec x \int_0^T dt  K^*(\vec x,t) = 1-S^*(T) $,
the function of Eq. \ref{Qkillingalone} reads
\begin{small}
\begin{eqnarray}
&& Q_T^{[  K^*(.,.)  ; S^*(T) ]}(\vec x,t)  
= \int_t^{T} dt_d  \int d^d \vec  x_d K^*(\vec x_d,t_d )
  \frac{P(\vec x_d,t_d \vert \vec  x,t) }{P(\vec x_d,t_d \vert \vec x_0,0)} 
 +S^*(T)  \left( \frac{S(T \vert \vec  x,t) }{ S(T \vert \vec x_0,0)} \right) 
\nonumber \\
&& = 
 \int_t^{T} dt_d  \int d^d \vec  x_d K^*(\vec x_d,t_d )
 \left( \frac{\sinh [\omega t_d]}{ \sinh [\omega (t_d-t)]} \right)^{\frac{d}{2}}
  e^{ 
   \displaystyle 
  \frac{\omega (\vec x_d^2 + \vec x_0^2) }{4 D \tanh [\omega t_d]}
  -\frac{\omega   (\vec x_d^2 + \vec x^2)  }{4 D \tanh [\omega (t_d-t)]}   
  +\frac{\omega  \vec x_d. \vec x}{2 D \sinh [\omega (t_d-t)]}
  - \frac{\omega  \vec x_d. \vec x_0}{2 D \sinh [\omega t_d]}
 }  
\nonumber \\
&&  +S^*(T)  
\frac{\cosh^{\frac{d}{2}} [\omega T]}{ \cosh^{\frac{d}{2}} [\omega (T-t)]}  
e^{ \displaystyle   \frac{\omega \tanh [\omega T]}{4 D }  \vec x_0^2 -  \frac{\omega \tanh [\omega (T-t)]}{4 D }  \vec x^2} 
 \label{Qkillingaloneoh}
\end{eqnarray}
\end{small}


\subsection{ Conditioning towards the surviving probability $ S^*(T) $ at time $T$ alone}

If one wishes to impose only the value $S^*(T)$ of the conditioned survival probability at time $T$,
 the function of Eq. \ref{Qsurvivingalone} reads
\begin{eqnarray}
&& Q_T^{[  S^*(T)]}(\vec x,t) 
 = \left[1- S^*(T) \right] \left( \frac{1-S(T \vert  x,t) }{1- S(T \vert \vec x_0,0)} \right) 
+ S^*(T) \left( \frac{ S(T \vert \vec  x,t) }{ S(T \vert \vec x_0,0)} \right) 
\nonumber \\
&& =
 \left[1- S^*(T) \right] 
\left( \frac{1-  \frac{1}{ \cosh^{\frac{d}{2}} [\omega (T-t)]}  e^{  -  \frac{\omega \tanh [\omega (T-t)]}{4 D }  \vec x^2} }
{1- \frac{1}{ \cosh^{\frac{d}{2}} [\omega T]} e^{  -  \frac{\omega \tanh [\omega T]}{4 D }  \vec x_0^2} 
} \right)
+ S^*(T) 
\frac{\cosh^{\frac{d}{2}} [\omega T]}{ \cosh^{\frac{d}{2}} [\omega (T-t)]}  
e^{  \frac{\omega \tanh [\omega T]}{4 D }  \vec x_0^2 -  \frac{\omega \tanh [\omega (T-t)]}{4 D }  \vec x^2}  
 \label{Qsurvivingaloneoh}
\end{eqnarray}


\subsubsection{ Special case : conditioning towards the survival probability $S^*(T)=1 $ 
at the finite horizon $T$ }

\begin{figure}[h]
\centering
\includegraphics[width=4.2in,height=3.2in]{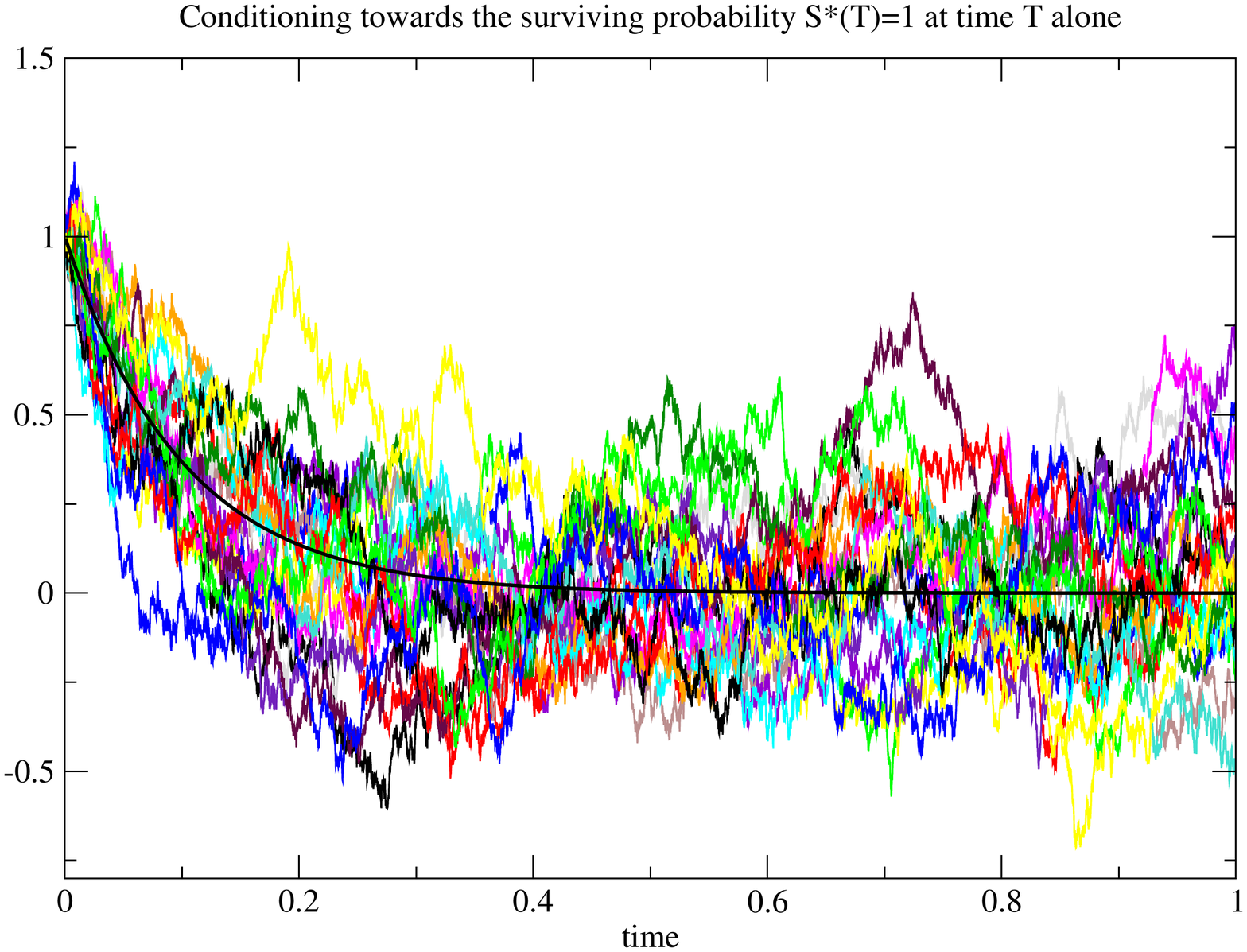}
\setlength{\abovecaptionskip}{15pt}  
\caption{A sample of $20$ diffusions satisfying the Ito stochastic differential equation Eq. \ref{Itodriftdoobfpstars1quadratic} with $\omega = 10$. Each color corresponds to the realization of one process. The thick black curve is the average profile of the stochastic process as given by equation \ref{averageItodriftohSurvival1delta}. The time step used in the discretization is $dt = 10^{-4}$.}
\label{fig2}
\end{figure}

For the special case where one wishes to impose the survival with probability unity $S^*(T)=1 $ 
at the finite horizon $T$, 
the conditioned killing rate of Eq. \ref{killingratestarsurvival} of course vanishes 
\begin{eqnarray}
k_T^*(\vec x,t) = 0
\label{killingratestarsurvivalstars1oh}
\end{eqnarray}
while the function of Eq. \ref{Qsurvivingaloneoh} reduces to 
\begin{eqnarray}
Q_T^{[  S^*(T)=1]}(\vec x,t) 
 =  \frac{\cosh^{\frac{d}{2}} [\omega T]}{ \cosh^{\frac{d}{2}} [\omega (T-t)]}  
e^{  \frac{\omega \tanh [\omega T]}{4 D }  \vec x_0^2 -  \frac{\omega \tanh [\omega (T-t)]}{4 D }  \vec x^2}  
 \label{Qsurvivingalonestars1oh}
\end{eqnarray}
The corresponding conditioned drift of Eq. \ref{driftdoobfp} reduces to
\begin{eqnarray}
 \vec \mu_T^*(\vec x,t)  =  2 D   \vec \nabla \ln Q_T^{[  S^*(T)=1]}(\vec x,t) 
 =  -  \omega \tanh [\omega (T-t)]  \vec x
\label{driftdoobfpstars1quadratic}
\end{eqnarray}
in agreement with \cite{Frydman}.
In the limit of the infinite horizon $T \to +\infty$, the conditioned drift
\begin{eqnarray}
 \vec \mu_{\infty}^*(\vec x)   =  -  \omega   \vec x
\label{mustarOU}
\end{eqnarray}
corresponds to the Ornstein-Uhlenbeck process. In one dimension, the conditioned process satisfies the Ito stochastic differential equation 
\begin{eqnarray}
 dX^*(t) =  - \omega \tanh [\omega (T-t)] X^*(t)  dt + dW(t)
 \label{Itodriftdoobfpstars1quadratic}
\end{eqnarray}
Figure \ref{fig2} shows a set of $20$ realizations of the process with parameter $\omega = 10$ as well as the mean trajectory $\langle X^*(t) \rangle$. This quantity is obtained by averaging the preceding equation over the realizations. Since $\langle dW(t) \rangle = 0$, we have
\begin{eqnarray}
 \frac{ d \langle X^*(t)\rangle}{dt} =  - \omega \tanh [\omega (T-t)]  \langle X^*(t)\rangle 
 \label{averageeqdiffItodriftdoobfpstars1quadratic}
\end{eqnarray}
Solving Eq. \ref{averageeqdiffItodriftdoobfpstars1quadratic} is straightforward and we get 
\begin{eqnarray}
 \langle X^*(t) \rangle =  x_0 \frac{\cosh(\omega (T-t))}{\cosh(\omega T)}
 \label{averageItodriftohSurvival1deltabis}
\end{eqnarray}

The conditioning towards $S^*(T)=1 $ is discussed in detail in \cite{Frydman} for 
the more general case of Gaussian diffusions with quadratic killing rates
with time-dependent parameters.

\subsubsection{ Special case : conditioning towards the survival probability $S^*(T)=0 $ 
at the finite horizon $T$ }

For the special case where one wishes to impose the survival with probability $S^*(T)=0 $ 
at the finite horizon $T$, 
 the function of Eq. \ref{Qsurvivingaloneoh} reduces to 
 \begin{eqnarray}
 Q_T^{[  S^*(T)=0]}(\vec x,t) =
 \frac{1-  \frac{1}{ \cosh^{\frac{d}{2}} [\omega (T-t)]}  e^{  -  \frac{\omega \tanh [\omega (T-t)]}{4 D }  \vec x^2} }
{1- \frac{1}{ \cosh^{\frac{d}{2}} [\omega T]} e^{  -  \frac{\omega \tanh [\omega T]}{4 D }  \vec x_0^2} 
 } 
 \label{Qsurvivingaloneohstars0}
\end{eqnarray}
The corresponding conditioned drift of Eq. \ref{driftdoobfp} reads
\begin{eqnarray}
 \vec \mu_T^*(\vec x,t)  =  2 D   \vec \nabla \ln Q_T^{[  S^*(T)=0]}(\vec x,t) 
 =    \frac{  \omega \tanh [\omega (T-t)] }
 { \cosh^{\frac{d}{2}} [\omega (T-t)]  e^{    \frac{\omega \tanh [\omega (T-t)]}{4 D }  \vec x^2} -1 } \ \vec x
\label{driftdoobfpstars1quadraticstars0}
\end{eqnarray}
while the conditioned killing rate of Eq. \ref{killingratestarsurvivalstar0} is given by
\begin{eqnarray}
k_T^*(\vec x,t)  
= \frac{ k (\vec x)   } { 1-S(T \vert \vec  x,t)} 
=  \frac{  \frac{\omega^2}{4D} \vec x^2 
  } {1-  \frac{1}{ \cosh^{\frac{d}{2}} [\omega (T-t)]}  e^{  -  \frac{\omega \tanh [\omega (T-t)]}{4 D }  \vec x^2} } 
\label{killingratestarsurvivalstar0oh}
\end{eqnarray}
Observe that when $\omega \to 0$ the conditioned process tends to a limit process whose parameters are given by 
\begin{eqnarray}
\lim_{\omega \to 0}  \vec \mu_T^*(\vec x,t)  = \frac{4 D \vec x}{\vec x^2 + D d(T-t)}
\label{driftdoobfpstars1quadraticstars0omega0}
\end{eqnarray}
and
\begin{eqnarray}
\lim_{\omega \to 0}  k_T^*(\vec x,t)  
=  \frac{\vec x^2} {(T-t) (\vec x^2 + D d(T-t))}
\label{killingratestarsurvivalstar0omega0}
\end{eqnarray}
Observe also that when $t \to T$, the killing rate  Eq. \ref{killingratestarsurvivalstar0oh} diverges as $\sim 1/(T-t)$, ensuring that the process cannot survive at times greater than $T$.
 
\begin{figure}[h]
\centering
\includegraphics[width=4.2in,height=3.2in]{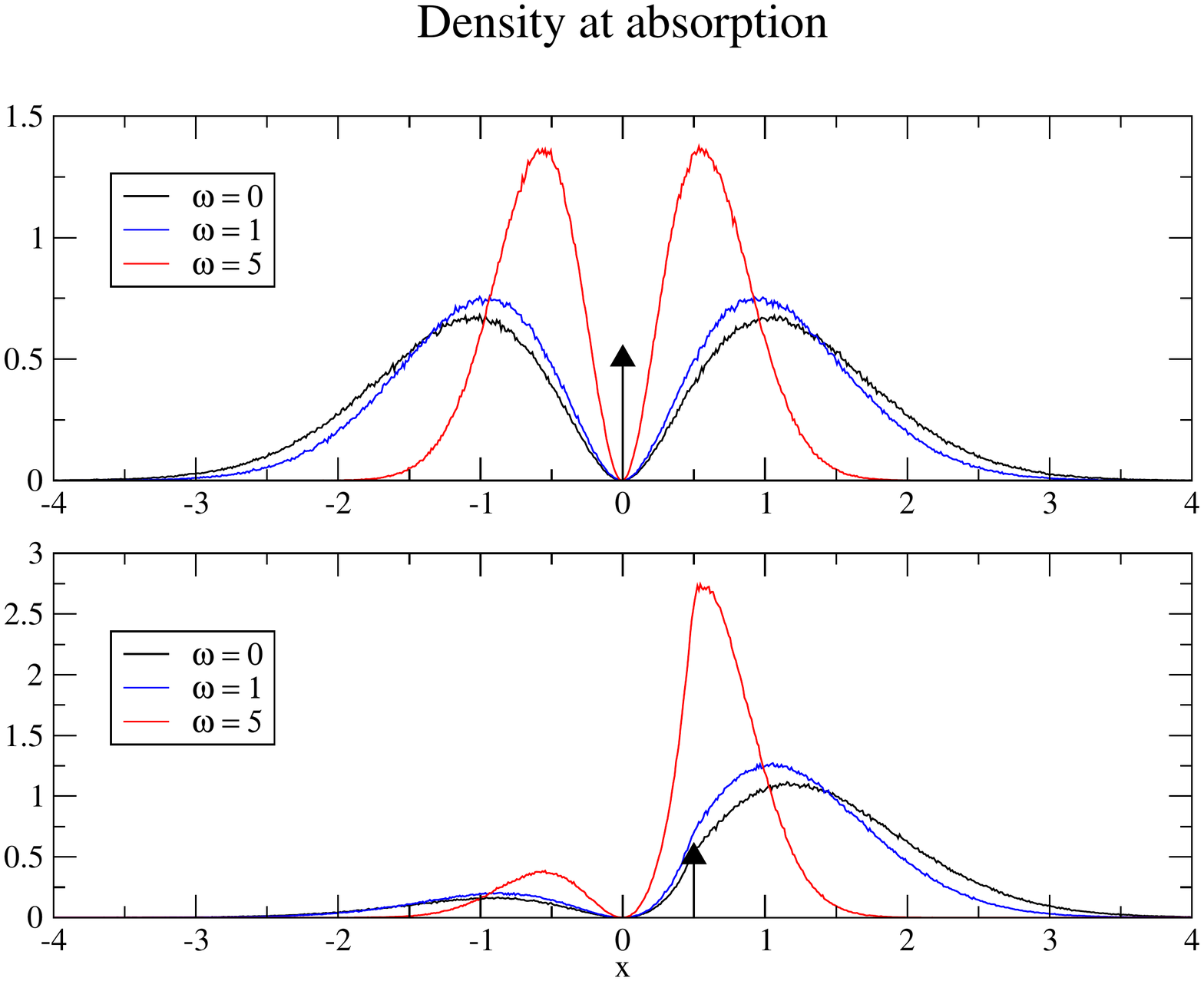}
\setlength{\abovecaptionskip}{15pt}  
\caption{Profile of the density at absorption for the process with drift and killing rate given by the Eqs. \ref{driftdoobfpstars1quadraticstars0} and  \ref{killingratestarsurvivalstar0oh} with different intensities of the killing rate. The arrow indicates the starting position of the process: upper figure $x_0 = 0$ (symmetrical case), lower figure $x_0 = 0.5$. Due to the killing rate $\sim x^2$, no absorption occurs at $x = 0$, as expected.}
\label{fig3}
\end{figure}



\section{Application to the Brownian with uniform drift and delta killing rate  }

\label{sec_delta}

In this section, the general construction is applied to the one-dimensional Brownian with uniform drift
when the killing rate is a delta function localized at the origin $x=0$.

\subsection{ Unconditioned process $  X(t)$ : Brownian with uniform drift $ \mu ( x)=  \mu \geq 0$
and delta killing rate $k ( x)=k\delta(x)$ }

The unconditioned process is the one-dimensional Brownian with uniform drift $\mu(x)=\mu \geq 0$, with uniform diffusion coefficient $D(x)=\frac{1}{2}$, while the killing rate is a delta function of amplitude $k>0$ localized at the origin $x=0$
\begin{eqnarray}
k(x)= k \delta(x)
\label{delta}
\end{eqnarray}
The amplitude $k \in ]0,+\infty[ $ allows to interpolate between the no-killing case $k =0$ and the 
absorbing condition at the origin that can be recovered in the limit of infinite amplitude $k \to +\infty$.


\subsubsection{ Reminder on the unconditioned propagator $P(  x,t \vert  x_0,t_0)$}

The forward dynamics of Eq. \ref{forward} for the propagator $P(  x,t \vert  x_0,t_0)$
\begin{eqnarray}
\partial_t P(  x,t \vert   x_0,t_0) 
= -      \mu \partial_x P(  x,t \vert   x_0,t_0) +  \frac{1}{2} \partial_x^2  P(  x,t \vert   x_0,t_0) 
-  k \delta(x) P(  0,t \vert   x_0,t_0) 
\label{forward1d}
\end{eqnarray}
can be translated for the time Laplace transform
\begin{eqnarray}
{\tilde P}_s (x \vert x_0) \equiv \int_{t_0}^{+\infty} dt e^{-s (t-t_0) } P(  x,t \vert   x_0,t_0) 
\label{laplace}
\end{eqnarray}
into
\begin{eqnarray}
- \delta(x-x_0) + s {\tilde P}_s (x \vert x_0)
= -      \mu \partial_x {\tilde P}_s (x \vert x_0) +  \frac{1}{2} \partial_x^2  {\tilde P}_s (x \vert x_0)
-  k \delta(x) {\tilde P}_s (0 \vert x_0)
\label{forward1dlaplace}
\end{eqnarray}

When there is no killing $k=0$, the free gaussian propagator 
\begin{eqnarray}
G(x,t \vert x_0,t_0) \equiv \frac{1}{\sqrt{2 \pi (t-t_0)}}  e^{- \frac{[x-x_0-\mu(t-t_0) ]^2}{2(t-t_0)}} 
=\frac{1}{\sqrt{2 \pi (t-t_0)}}  
e^{- \frac{(x-x_0)^2}{2(t-t_0)} 
+\mu (x-x_0)
- \frac{\mu^2 }{2}(t-t_0)
}
\label{free}
\end{eqnarray}
has for Laplace transform
\begin{eqnarray}
{\tilde G}_s (x \vert x_0) && \equiv \int_{t_0}^{+\infty} dt e^{-s (t-t_0) } G(  x,t \vert   x_0,t_0) 
= \frac{ e^{\mu (x-x_0)} } { \sqrt{2 \pi }} 
\int_{0}^{+\infty} d\tau \tau^{-\frac{1}{2}} e^{-\left(s + \frac{\mu^2}{2} \right) \tau }
e^{- \frac{(x-x_0)^2}{2\tau} }
\nonumber \\
&& = \frac{ e^{\mu (x-x_0) - \sqrt{\mu^2+2 s } \vert x-x_0 \vert } } { \sqrt{\mu^2+2 s }} 
\label{laplacefree}
\end{eqnarray}
that solves Eq. \ref{forward1dlaplace} for $k=0$.
For $k \ne 0$, the solution ${\tilde P}_s (x \vert x_0) $ can be obtained from this free solution ${\tilde G}_s (x \vert x_0) $ via the resummation of the Dyson perturbative series in the parameter $k$
\begin{eqnarray}
 {\tilde P}_s (x \vert x_0)
&& = {\tilde G}_s (x \vert x_0) -k {\tilde G}_s (x \vert 0)  {\tilde G}_s (0 \vert x_0)
+ k^2 {\tilde G}_s (x \vert 0)  {\tilde G}_s (0 \vert 0)   {\tilde G}_s (0 \vert x_0)
\nonumber \\
&&-k^3 {\tilde G}_s (x \vert 0)  {\tilde G}_s (0 \vert 0)   {\tilde G}_s (0 \vert 0)  {\tilde G}_s (0 \vert x_0) +...
\nonumber \\
&& =  {\tilde G}_s (x \vert x_0) - {\tilde G}_s (x \vert 0)  \frac{ k   }{ 1+ k {\tilde G}_s (0 \vert 0)}{\tilde G}_s (0 \vert x_0)
\label{resum}
\end{eqnarray}
and thus reads
using the explicit expression of Eq. \ref{laplacefree}
\begin{eqnarray}
 {\tilde P}_s (x \vert x_0)
&& =   \frac{ e^{\mu (x-x_0) - \sqrt{\mu^2+2 s } \vert x-x_0 \vert } } { \sqrt{\mu^2+2 s }} 
-  \frac{ e^{\mu x - \sqrt{\mu^2+2 s } \vert x \vert } } { \sqrt{\mu^2+2 s }} 
\left(   \frac{ k   }{ 1+ \frac{k} { \sqrt{\mu^2+2 s }} }  \right)
 \frac{ e^{- \mu x_0 - \sqrt{\mu^2+2 s } \vert x_0 \vert } } { \sqrt{\mu^2+2 s }} 
 \nonumber \\
&& 
= e^{\mu (x-x_0)} \left[ \frac{ e^{ - \sqrt{\mu^2+2 s } \vert x-x_0 \vert } } { \sqrt{\mu^2+2 s }} 
-  \frac{ e^{ - \sqrt{\mu^2+2 s } (\vert x \vert +\vert x_0 \vert ) } } { \sqrt{\mu^2+2 s }} 
\left(   \frac{ k   }{ k + \sqrt{\mu^2+2 s } }  \right)
 \right]
\label{resumexpli}
\end{eqnarray}
It is useful to write the last factor as an integral over another variable $z $
\begin{eqnarray}
 {\tilde P}_s (x \vert x_0)
&&  = e^{\mu (x-x_0)} \left[ \frac{ e^{ - \sqrt{\mu^2+2 s } \vert x-x_0 \vert } } { \sqrt{\mu^2+2 s }} 
- k \frac{ e^{ - \sqrt{\mu^2+2 s } (\vert x \vert +\vert x_0 \vert ) } } { \sqrt{\mu^2+2 s }} 
\int_0^{+\infty} dz e^{ - \left( k + \sqrt{\mu^2+2 s } \right) z}
 \right]
   \nonumber \\
&& = e^{\mu (x-x_0)} \left[ \frac{ e^{ - \sqrt{\mu^2+2 s } \vert x-x_0 \vert } } { \sqrt{\mu^2+2 s }} 
- k \int_0^{+\infty} dz e^{- k z} 
\frac{ e^{ - \sqrt{\mu^2+2 s } (\vert x \vert +\vert x_0 \vert +z) } } { \sqrt{\mu^2+2 s }} 
 \right]
\label{resumexpliinteg}
\end{eqnarray}
in order to rewrite the second contribution of Eq. \ref{resumexpli}
in terms of the free propagator of Eq. \ref{laplacefree}
at another argument
\begin{eqnarray}
{\tilde G}_s (\vert x \vert +\vert x_0 \vert +z \vert 0)  = 
\frac{ e^{ (\mu  - \sqrt{\mu^2+2 s }) (\vert x \vert +\vert x_0 \vert +z) } } { \sqrt{\mu^2+2 s }} 
\label{laplacefreeanother}
\end{eqnarray}
to obtain
\begin{eqnarray}
 {\tilde P}_s (x \vert x_0)
 = {\tilde G}_s (x \vert x_0)
- k \int_0^{+\infty} dz e^{- k z} e^{\mu (x-x_0)} e^{ - \mu (\vert x \vert +\vert x_0 \vert +z)}
{\tilde G}_s (\vert x \vert +\vert x_0 \vert +z \vert 0)
\label{resumexpliG}
\end{eqnarray}
The Laplace inversion is now straighforward
and one obtains that the propagator $P(  x,t \vert   x_0,t_0) $
reads  in terms of the Gaussian propagator $G(  x,t \vert   x_0,t_0) $ of Eq. \ref{free}
\begin{eqnarray}
&& P(  x,t \vert   x_0,t_0)
 = G(  x,t \vert   x_0,t_0)
- k e^{  \mu (x- \vert x \vert) -\mu (x_0+\vert x_0 \vert)} 
\int_0^{+\infty} dz e^{- (k +\mu) z  }
G (\vert x \vert +\vert x_0 \vert +z , t-t_0 \vert 0,0)
\nonumber \\
&& = \frac{e^{- \frac{(x-x_0)^2}{2(t-t_0)} 
+\mu (x-x_0)
- \frac{\mu^2 }{2}(t-t_0)}}{\sqrt{2 \pi (t-t_0)}}  
- k e^{  \mu (x- \vert x \vert) -\mu (x_0+\vert x_0 \vert)} 
\int_0^{+\infty} dz e^{- (k +\mu) z  }
\frac{e^{- \frac{(\vert x \vert +\vert x_0 \vert +z)^2}{2(t-t_0)} 
+\mu (\vert x \vert +\vert x_0 \vert +z)
- \frac{\mu^2 }{2}(t-t_0)}}{\sqrt{2 \pi (t-t_0)}}  
\nonumber \\
&& = \frac{e^{\mu (x-x_0)- \frac{\mu^2 }{2}(t-t_0)}} {\sqrt{2 \pi (t-t_0)}}  
\left[ e^{- \frac{(x-x_0)^2}{2(t-t_0)} }
- k 
\int_0^{+\infty} dz e^{- k  z - \frac{(\vert x \vert +\vert x_0 \vert +z)^2}{2(t-t_0)} }  \right] 
\nonumber \\
&& = \frac{e^{\mu (x-x_0)- \frac{\mu^2 }{2}(t-t_0)}} {\sqrt{2 \pi (t-t_0)}}  
\left[ e^{- \frac{(x-x_0)^2}{2(t-t_0)} } -  k \sqrt{\frac{\pi (t-t_0)}{2}} e^{\frac{k}{2} (2 (\left| x\right| +\left|x_0\right|) +k (t-t_0))} \erfc\left(\frac{\left| x\right| +\left|x_0\right| +k (t-t_0)}{\sqrt{2(t-t_0)}}\right)\right] 
\label{propagatordelta}
\end{eqnarray}

with the complementary Error function $\erfc(x) = 1 - \erf(x) = 1 - \frac{2}{\sqrt{\pi}} \int_0^{x} du \, e^{- u^2}$.


\subsubsection{ Survival probability $ S(T \vert   x,t)$  }

The survival probability of Eq. \ref{survival} can be obtained via the integration of Eq. \ref{propagatordelta} over $x_T$
\begin{eqnarray}
S(T \vert   x,t)  && = \int_{-\infty}^{+\infty} d x_T P(  x_T,T \vert   x,t) 
= \int_{-\infty}^{+\infty} d x_T 
 \frac{e^{\mu (x_T-x)- \frac{\mu^2 }{2}(T-t)}} {\sqrt{2 \pi (T-t)}}  
\left[ e^{- \frac{(x_T-x)^2}{2(T-t)} }
- k 
\int_0^{+\infty} dz e^{- k  z - \frac{(\vert x_T \vert +\vert x \vert +z)^2}{2(T-t)} }  \right]
\nonumber \\
&& = 1 
- k  \frac{e^{- \mu x- \frac{\mu^2 }{2}(T-t)}} {\sqrt{2 \pi (T-t)}}  
\int_{-\infty}^{+\infty} d x_T e^{ \mu x_T} 
\int_0^{+\infty} dz e^{- k  z 
- \frac{x_T ^2}{2(T-t)} 
- \frac{2 \vert x_T \vert (\vert x \vert +z)}{2(T-t)} 
- \frac{(\vert x \vert +z)^2}{2(T-t)} 
}  
\nonumber \\
&& = 1 
- \frac{k}{2}  e^{- \mu x + k \left| x\right| + \frac{1}{2} (k^2-\mu^2 ) (T-t) }     
 \int_{-\infty}^{+\infty} d x_T  e^{\mu x_T + k \left| x_T\right|}
\erfc\left(\frac{\left| x\right| +\left| x_T\right| + k (T-t)}{\sqrt{2(T-t)}}\right)
\label{survivaldelta}
\end{eqnarray}
It is thus more convenient to return to the propagator Laplace transform of Eq. \ref{resumexpli}
where the integration over $x_T$ can be explicitly performed 
in order to obtain the following Laplace transform $ {\tilde S}_s ( x) $ of the survival propability $S(T \vert   x,t)  $  
\begin{eqnarray}
&& {\tilde S}_s ( x)  \equiv \int_{t}^{+\infty} dT e^{-s (T-t) } S(T \vert   x,t) 
=  \int_{-\infty}^{+\infty} d x_T {\tilde P}_s (  x_T \vert   x) 
\nonumber \\
&& = \frac{ e^{-\mu x}  } { \sqrt{\mu^2+2 s }} 
\int_{-\infty}^{+\infty} d x_T
e^{\mu x_T - \sqrt{\mu^2+2 s } \vert x_T-x \vert } 
-  \frac{ k e^{ - \mu x - \sqrt{\mu^2+2 s } \vert x \vert   }   }{ \left( k + \sqrt{\mu^2+2 s } \right) \sqrt{\mu^2+2 s }} 
 \int_{-\infty}^{+\infty} d x_T
e^{\mu x_T - \sqrt{\mu^2+2 s } \vert x_T \vert  }
\nonumber \\
&& = \frac{ e^{-\mu x}  } { \sqrt{\mu^2+2 s }} 
\left[ e^{ - \sqrt{\mu^2+2 s } x} \int_{-\infty}^{x} d x_T e^{(\mu + \sqrt{\mu^2+2 s } ) x_T } 
+e^{  \sqrt{\mu^2+2 s } x} \int_{x}^{+\infty} d x_T e^{ (\mu - \sqrt{\mu^2+2 s } ) x_T } 
\right] 
\nonumber \\
&& -  \frac{ k e^{ - \mu x - \sqrt{\mu^2+2 s } \vert x \vert   }   }{ \left( k + \sqrt{\mu^2+2 s } \right) \sqrt{\mu^2+2 s }} 
\left[  \int_{-\infty}^{0} d x_T e^{ (\mu  + \sqrt{\mu^2+2 s } ) x_T   }  
+ \int_{0}^{+\infty} d x_T e^{ ( \mu  - \sqrt{\mu^2+2 s } ) x_T   }  
\right] 
\nonumber \\
&& = \frac{ e^{-\mu x}  } { \sqrt{\mu^2+2 s }} 
\left[ e^{ - \sqrt{\mu^2+2 s } x} \frac{ e^{(\mu + \sqrt{\mu^2+2 s } ) x } }{  \sqrt{\mu^2+2 s } +\mu } 
+e^{  \sqrt{\mu^2+2 s } x}  \frac{ e^{(\mu - \sqrt{\mu^2+2 s } ) x } }{  \sqrt{\mu^2+2 s } -\mu } 
\right] 
\nonumber \\
&& -  \frac{ k e^{ - \mu x - \sqrt{\mu^2+2 s } \vert x \vert   }   }{ \left( k + \sqrt{\mu^2+2 s } \right) \sqrt{\mu^2+2 s }} 
\left[  \frac{1} {  \sqrt{\mu^2+2 s } +\mu } 
+ \frac{1} {  \sqrt{\mu^2+2 s } -\mu } 
\right] 
\nonumber \\
&& = \frac{ 1 } { s}  \left[ 1 -  \frac{ k e^{ - \mu x - \sqrt{\mu^2+2 s } \vert x \vert   }   }{  \left( k + \sqrt{\mu^2+2 s } \right)  } \right]
 \label{laplaceS}
\end{eqnarray}
This expression allows to obtain the unconditioned survival probability for large time $T \to +\infty$
as discussed in the two following subsections.


\subsubsection{ Unconditioned forever-survival probability $ S(\infty \vert   x)$}

In Eq. \ref{laplaceS}, the coefficient of $\frac{1}{s}$ as $s \to 0$ corresponds to the forever survival probability $S(\infty \vert x) $
\begin{eqnarray}
 {\tilde S}_s ( x) \opsimeq_{s \to 0} \frac{ S(\infty \vert x) }{s } 
 \label{coef1surs}
\end{eqnarray}
so one obtains using the positivity of the drift $\mu \geq 0$
\begin{eqnarray}
 S(\infty \vert x) = 1 -  \frac{ k e^{ - \mu (x +  \vert  x \vert )  }   }{   k +  \mu   } 
 = 
 \left\lbrace
  \begin{array}{lll}
      1 -  \frac{ k e^{ - 2 \mu x }   }{   k +  \mu   } 
    &~~\mathrm{if~~} x \ge 0
    \\
      \frac{ \mu    }{   k +  \mu   } 
    &~~\mathrm{if~~} x \le 0  
  \end{array}
\right.
 \label{survivalinftydelta}
\end{eqnarray}
This result can also be found directly by solving the backward dynamics of Eq. \ref{Sbackwardinfty}
\begin{eqnarray}
0  = {\cal F} S(\infty \vert   x) 
=  \ \mu \partial_x S(\infty \vert  x)  + \frac{1}{2}  \partial_x^2 S(\infty \vert   x) - k \delta(x) S(\infty \vert  0) 
\label{Sbackwardinftydelta}
\end{eqnarray}

The physical meaning of the result of Eq. \ref{survivalinftydelta} can be understood as follows :

(i) for vanishing drift $\mu=0$, the Brownian motion returns an infinite number of times at the origin $x=0$
where it will be eventually killed, so the forever survival probability vanishes for any starting point $x$
\begin{eqnarray}
 S(\infty \vert x) = 0 \ \ { \rm for } \ \ \mu=0
 \label{survivalinftymuzero}
\end{eqnarray}

(ii) for strictly positive drift $\mu>0$, the Brownian motion returns only a finite number of times to the origin $x=0$
before flowing towards $(x \to +\infty)$, so the forever-survival probability $S(\infty \vert x) $
remains finite for any starting point $x$. When the initial condition is negative $x<0$, the particle has to cross the origin at least once,
so the survival probability is simply equal to the survival when one starts from the origin :
\begin{eqnarray}
 S(\infty \vert x<0) = S(\infty \vert x=0) \ \ { \rm for } \ \ \mu>0
 \label{survivalinftymuposxneg}
\end{eqnarray}
When the initial condition is positive $x>0$, 
the particle can escape towards $(+\infty)$ without ever touching the origin $x=0$ 
with the probability 
\begin{eqnarray}
 p_{escape}(x)= 1- e^{-2 \mu x } 
 \label{pescape}
\end{eqnarray}
while it will touch the origin $x=0$ at least once with the complementary probability $(1-p_{escape}(x))=  e^{-2 \mu x } $. As a consequence,
the result of the survival probability in Eq. \ref{survivalinftydelta}
can be understood as
\begin{eqnarray}
 S(\infty \vert x>0) = p_{escape}(x) + \left[ 1- p_{escape}(x) \right] S(\infty \vert x=0) \ \ { \rm for } \ \ \mu>0
 \label{survivalinftymuposxpos}
\end{eqnarray}


\subsubsection{ Unconditioned survival probability $ S(T \vert   .,.)$ for large $T$ when the drift vanishes $\mu=0$}

When the drift vanishes $\mu=0$,
the unconditioned survival probability of Eq. \ref{survivalinftymuzero}
vanishes $S(+\infty \vert .) =0$. In order to obtain the asymptotic behavior of $S(T \vert x, t ) $ for large time $T$,
one should compute the leading order of the Laplace transform of Eq. \ref{laplaceS}
for small $s \to 0^+$ 
\begin{eqnarray}
 {\tilde S}^{[\mu=0]}_s ( x) && = \frac{ 1 } { s} \left[ 1 -  \frac{ k e^{  - \sqrt{2 s } \vert x \vert   }   }{  \left( k + \sqrt{2 s } \right)  } \right]
 = \frac{1}{s} \left[ 1- \frac{1}{1 + \frac{\sqrt{2 s }}{k} } e^{  - \sqrt{2 s } \vert x \vert   }\right]
  = \frac{1}{s} \left[ 1- \left( 1 - \frac{\sqrt{2 s }}{k}  +O(s) \right) 
 \left( 1  - \sqrt{2 s } \vert x \vert   + O(s)\right) \right]
  \nonumber \\ &&
   = \sqrt{ \frac{2}{s} } \left[  \frac{1}{k}  +  \vert x \vert  \right] +    O(s^0)
 \label{laplaceSmuzero}
\end{eqnarray}
The Laplace inversion yields that the leading order of the survival probability $S(T \vert  x,t) $ 
when the time interval $(T-t)$ is large is given by
\begin{eqnarray}
S(T \vert  x,t) \opsimeq_{(T-t) \to +\infty} \sqrt{ \frac{2}{\pi (T-t) } } \left[  \frac{1}{k}  +  \vert x \vert  \right]
 \label{survivalmuzerolargeT}
\end{eqnarray}
that generalizes the survival probability for an absorbing condition at the origin that corresponds to $k \to +\infty$.


\subsection{ Full conditioning constraints $\left[ P^*(.,T) ; K^*(.)  \right]  $ associated to the finite horizon $T$ }

Using the propagator of Eq. \ref{propagatordelta},
the function $Q_T(x,t) $ of Eq. \ref{Qdef} reads
\begin{eqnarray}
  \label{Qdelta}
&& Q_T(x,t)   = 
 \int_t^{T} dt_d   K^*(t_d )  \frac{P(0,t_d \vert  x,t) }{P(0,t_d \vert  x_0,0)} 
 +  \int_{-\infty}^{+\infty} d y P^*( y,T ) \frac{ P( y,T \vert  x,t)  }{P( y,T \vert  x_0,0) }
 \nonumber \\
 && =
  \int_t^{T} dt_d   K^*(t_d )   
  \sqrt{ \frac{t_d}{t_d-t} }   e^{\mu (x_0-x)+ \frac{\mu^2 }{2}t}  
\left[  \frac{
 e^{- \frac{x^2}{2(t_d-t)} }
- k \int_0^{+\infty} dz e^{- k  z - \frac{(\vert x \vert +z)^2}{2(t_d-t)} } 
}
{ e^{- \frac{x_0^2}{2t_d} }
- k \int_0^{+\infty} dz_0 e^{- k  z_0 - \frac{(\vert x_0 \vert +z_0)^2}{2t_d} } 
} \right] 
 \nonumber \\
 && +  \int_{-\infty}^{+\infty} d y P^*( y,T ) 
\sqrt{ \frac{T}{T-t} }   e^{\mu (x_0-x)+ \frac{\mu^2 }{2}t}  
\left[  \frac{
 e^{- \frac{(y-x)^2}{2(T-t)} }
- k \int_0^{+\infty} dz e^{- k  z - \frac{(\vert y \vert +\vert x \vert +z)^2}{2(T-t)} } 
}
{ e^{- \frac{(y-x_0)^2}{2T} }
- k \int_0^{+\infty} dz_0 e^{- k  z_0 - \frac{(\vert y \vert +\vert x_0 \vert +z_0)^2}{2T} } 
} \right]
 \nonumber \\
 && =
  \int_t^{T} dt_d   K^*(t_d )   
  \sqrt{ \frac{t_d}{t_d-t} }   e^{\mu (x_0-x)+ \frac{\mu^2 }{2}t}  
\left[  \frac{
 e^{- \frac{x^2}{2(t_d-t)} }
- k \sqrt{\frac{\pi(t_d-t)}{2}} e^{\frac{1}{2} k (2 \left| x\right| +k (t_d-t))} \erfc \left( \frac{\left| x\right|+ k (t_d-t)}{\sqrt{2(t_d-t)}} \right)
}
{ e^{- \frac{x_0^2}{2t_d} }
- k \sqrt{\frac{\pi t_d}{2}} e^{\frac{1}{2} k (2 \left| x_0\right| +k t_d)} \erfc \left( \frac{\left| x_0\right|+ k t_d}{\sqrt{2 t_d}} \right)
} \right] 
  \\
 && +  \int_{-\infty}^{+\infty} d y P^*( y,T ) 
\sqrt{ \frac{T}{T-t} }   e^{\mu (x_0-x)+ \frac{\mu^2 }{2}t}  
\left[  \frac{
 e^{- \frac{(y-x)^2}{2(T-t)} }
- k \sqrt{\frac{\pi(T-t)}{2}} e^{\frac{1}{2} k (2 (\left| x\right| + \left| y\right|) + k (T-t))} \erfc \left( \frac{\left| x\right|+ \left| y\right| + k (T-t)}{\sqrt{2(T-t)}} \right) 
}
{ e^{- \frac{(y-x_0)^2}{2T} }
- k \sqrt{\frac{\pi T}{2}} e^{\frac{1}{2} k (2 (\left| x_0\right| + \left| y\right|) + k T)} \erfc \left( \frac{\left| x_0\right|+ \left| y\right| + k T}{\sqrt{2 T}} \right)
} \right] \nonumber
\end{eqnarray}


\subsubsection*{ Example of the bridge without being killed : $P^*( y,T )=\delta( y- y_*)   $  }

For the case where one imposes the full survival at time $T$ at the single position $ y_* $ 
\begin{eqnarray}
P^*( y,T ) && =\delta( y- y_*)
\nonumber \\
K^*(t_d ) && =0
 \label{pspatial1deltabis}
\end{eqnarray}
 Eq. \ref{Qdelta} reduces to
\begin{eqnarray}
 Q_T(x,t)  && = 
\sqrt{ \frac{T}{T-t} }   e^{\mu (x_0-x)+ \frac{\mu^2 }{2}t}  
\left[  \frac{
 e^{- \frac{(y_*-x)^2}{2(T-t)} }
- k \int_0^{+\infty} dz e^{- k  z - \frac{(\vert y_* \vert +\vert x \vert +z)^2}{2(T-t)} } 
}
{ e^{- \frac{(y_*-x_0)^2}{2T} }
- k \int_0^{+\infty} dz_0 e^{- k  z_0 - \frac{(\vert y_* \vert +\vert x_0 \vert +z_0)^2}{2T} } 
} \right] \nonumber \\
           && = 
\sqrt{ \frac{T}{T-t} }   e^{\mu (x_0-x)+ \frac{\mu^2 }{2}t}  
\left[  \frac{
 e^{- \frac{(y_*-x)^2}{2(T-t)} }
- k \sqrt{\frac{\pi}{2(T-t)}} e^{\frac{1}{2} k (2 (\vert y_* \vert +\vert x \vert)+k (T-t))} \erfc\left(\frac{\vert y_* \vert +\vert x \vert + k (T-t)}{\sqrt{2(T-t)}}\right)
}
{ e^{- \frac{(y_*-x_0)^2}{2T} }
- k \sqrt{\frac{\pi}{2 T}} e^{\frac{1}{2} k (2 (\vert y_* \vert +\vert x_0 \vert)+k T)} \erfc\left(\frac{\vert y_* \vert +\vert x_0 \vert + k T}{\sqrt{2 T}}\right) 
} \right] 
  \label{Qdeltabridge}
\end{eqnarray}

The corresponding conditioned drift of Eq. \ref{driftdoobfp} reads
\begin{eqnarray}
 \mu_T^*( x,t)  && = \mu + \partial_x  \ln Q_T (x,t)
=    \partial_x  \ln \left[ 
e^{- \frac{(y_*-x)^2}{2(T-t)} }
- k \int_0^{+\infty} dz e^{- k  z - \frac{(\vert y_* \vert +\vert x \vert +z)^2}{2(T-t)} } 
\right]
\nonumber \\
 && 
 = \frac{ \frac{y_*-x}{T-t} e^{- \frac{(y_*-x)^2}{2(T-t)} }
+ k \, {\rm sgn}(x) \int_0^{+\infty} dz \frac{\vert y_* \vert +\vert x \vert +z}{T-t } e^{- k  z - \frac{(\vert y_* \vert +\vert x \vert +z)^2}{2(T-t)} }}
 {e^{- \frac{(y_*-x)^2}{2(T-t)} }
- k \int_0^{+\infty} dz e^{- k  z - \frac{(\vert y_* \vert +\vert x \vert +z)^2}{2(T-t)} }}
 \\
 && 
 = \frac{ \frac{y_*-x}{T-t} e^{- \frac{(y_*-x)^2}{2(T-t)} }
+ k \, {\rm sgn}(x) \left[
  e^{-\frac{(\vert y_* \vert +\vert x \vert)^2}{2(T-t)}} - k \sqrt{\frac{\pi (T-t) }{2}}  e^{\frac{1}{2} k (2 (\vert y_* \vert +\vert x \vert)+k (T-t))} \erfc\left(\frac{\vert y_* \vert +\vert x \vert+k (T-t)}{ \sqrt{2(T-t)}}\right) \right]
 }
 {e^{- \frac{(y_*-x)^2}{2(T-t)} }
- k \sqrt{\frac{\pi}{2(T-t)}} e^{\frac{1}{2} k (2 (\vert y_* \vert +\vert x \vert)+k (T-t))} \erfc\left(\frac{\vert y_* \vert +\vert x \vert + k (T-t)}{\sqrt{2(T-t)}}\right) } \nonumber
    \label{driftohSurvivaldeltabridge}
\end{eqnarray}


\subsection{ Conditioning towards the surviving probability $ S^*(\infty) $ at the infinite horizon $T=+\infty$}

Here one needs to distinguich whether the unconditioned drift $\mu$ vanishes or not.


\subsubsection{ Case $\mu=0$ where the unconditioned
forever-survival probability vanishes $S(+\infty \vert .) =0$   }

For vanishing drift $\mu=0$,
the unconditioned survival probability of Eq. \ref{survivalinftymuzero}
vanishes $S(+\infty \vert .) =0$,
so one should use Eq. \ref{Qsurvivingaloneinftyzero} with the asymptotic behavior of Eq. \ref{survivalmuzerolargeT}
to obtain
\begin{eqnarray}
Q_{\infty}^{[  S^*(\infty)]}( x,t) 
&& =  [1-S^*(\infty)]
 + S^*(\infty) \lim_{T \to +\infty} \left( \frac{S(T \vert  x,t) }{ S(T \vert  x_0,0)} \right)
\nonumber \\
&&  =  [1-S^*(\infty)] + S^*(\infty) \frac{  \frac{1}{k}  +  \vert x \vert }{  \frac{1}{k}  +  \vert x_0 \vert } 
 \label{Qsurvivingaloneinftyzeromuzero}
\end{eqnarray}
The conditioned drift 
of Eq. \ref{driftdoobfp} reads
\begin{eqnarray}
 \mu_{\infty}^*( x)   =  \partial_x  \ln Q_{\infty}^{[  S^*(\infty)]}( x)
=  {\rm sgn}(x)
 \frac{  S^*(\infty)  }{\frac{1}{k} + \vert x_0 \vert (1-S^*(\infty))  + \vert x \vert S^*(\infty)  }
 \label{driftstarsurvivalinftydeltamuzero}
\end{eqnarray}
 while the conditioned killing rate of Eq. \ref{killingratestarsurvivalinftyfinite} is given by
\begin{eqnarray}
k_{\infty}^*( x)  =  \frac{ [ 1- S^*(\infty)]    }
{ Q_{\infty}^{[  S^*(\infty) ] }(0) } k \delta( x)
 = \frac{ [ 1- S^*(\infty) ]    }
{  [1-S^*(\infty)] + S^*(\infty) \frac{  1  }{  1  + k \vert x_0 \vert }  } k \delta( x)
 \label{killingratestarsurvivalinftyfinitedelta}
\end{eqnarray}

For the special case where the conditioning is towards the forever-survival $S^*(\infty)=1$, 
the conditioned killing rate of course vanishes $k^*(x)=0$, while
the drift of Eq. \ref{driftstarsurvivalinftydeltamuzero}
reduces to
\begin{eqnarray}
 \mu_{\infty}^*( x)  
=  {\rm sgn}(x)
 \frac{  1  }{\frac{1}{k}  + \vert x \vert  }
 \label{driftstarsurvivalinftydeltamuzeros1}
\end{eqnarray}
For $x>0$, one recovers the Bessel drift $\mu_{Bessel}(x)=\frac{1}{x} $ in the limit $k \to +\infty$.

\begin{figure}[h]
\centering
\includegraphics[width=4.8in,height=4.2in]{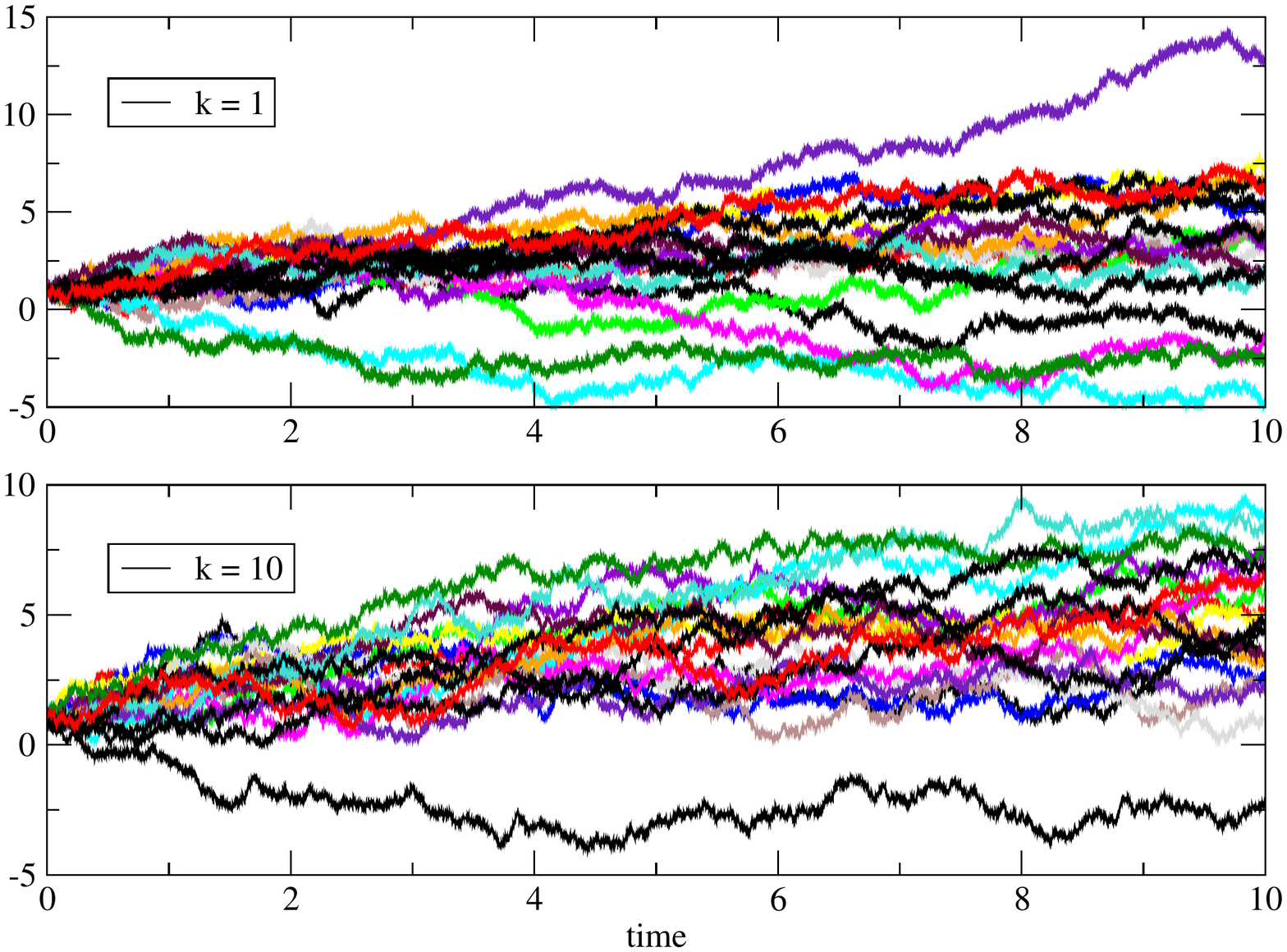}
\setlength{\abovecaptionskip}{15pt}  
\caption{A sample of $20$ diffusions satisfying the drift given by Eq. \ref{driftstarsurvivalinftydeltamuzeros1} for two different values of the absorbing parameter $k$. Each color corresponds to the realization of one process. The time step used in the discretization is $dt = 10^{-4}$. For a parameter $k$ close to zero the process behaves like a free Brownian motion (top figure) then as $k$ increases the process behaves more and more like a Bessel process (bottom figure).}
\label{fig4}
\end{figure}


\subsubsection{ Case $\mu>0$ where the unconditioned
forever-survival probability is finite $S(+\infty \vert .) >0$   }

For strictly positive drift $\mu>0$,
the unconditioned survival probability of Eq. \ref{survivalinftydelta}
remains finite $S(+\infty \vert .) >0$,
so one should use Eq. \ref{Qsurvivingaloneinftyfinite} with Eq. \ref{survivalinftydelta}
\begin{eqnarray}
Q_{\infty}^{[  S^*(\infty)]}( x) 
&& =  (1-S^*(\infty))  \left( \frac{1-S(\infty \vert  x) }{1- S(\infty \vert  x_0)} \right)
 + S^*(\infty)  \left( \frac{S(\infty \vert  x) }{ S(\infty \vert  x_0)} \right)
 \nonumber \\
 && =
  (1-S^*(\infty)) e^{ \mu (x_0 +  \vert  x_0 \vert ) - \mu (x +  \vert  x \vert )} 
 + S^*(\infty) 
  \left[ \frac{  \mu +k (1 -  e^{ - \mu (x +  \vert  x \vert )  } )   }
 { \mu+k(1 -  e^{ - \mu (x_0 +  \vert  x_0 \vert )  } )   } 
 \right]
 \label{Qsurvivingaloneinftyfinitedelta}
\end{eqnarray}
The conditioned drift 
of Eq. \ref{driftdoobfp} reads
\begin{eqnarray}
 \mu_{\infty}^*( x)  && = \mu + \partial_x  \ln Q_{\infty}^{[  S^*(\infty)]}( x)
 \nonumber \\
&&= \mu - \mu (1+ {\rm sgn}(x) ) 
\frac{(1-S^*(\infty)) e^{ \mu (x_0 +  \vert  x_0 \vert ) - \mu (x +  \vert  x \vert )} 
 + S^*(\infty) 
  \left[ \frac{ -k   e^{ - \mu (x +  \vert  x \vert )  }    }
 { \mu+k(1 -  e^{ - \mu (x_0 +  \vert  x_0 \vert )  } )   } 
 \right]}
{(1-S^*(\infty)) e^{ \mu (x_0 +  \vert  x_0 \vert ) - \mu (x +  \vert  x \vert )} 
 + S^*(\infty) 
  \left[ \frac{  \mu +k (1 -  e^{ - \mu (x +  \vert  x \vert )  } )   }
 { \mu+k(1 -  e^{ - \mu (x_0 +  \vert  x_0 \vert )  } )   } 
 \right]}
  \nonumber \\
&&=
  \left\lbrace
  \begin{array}{lll}
    \mu  
    &~~\mathrm{if~~} x < 0
    \\
 \mu - 2 \mu 
\frac{(1-S^*(\infty)) e^{ \mu (x_0 +  \vert  x_0 \vert )   } 
 + S^*(\infty) 
  \left[ \frac{ -k       }
 { \mu+k(1 -  e^{ - \mu (x_0 +  \vert  x_0 \vert )  } )   } 
 \right]}
{(1-S^*(\infty)) e^{ \mu (x_0 +  \vert  x_0 \vert )  } 
 + S^*(\infty) 
  \left[ \frac{  (\mu +k ) e^{  2 \mu x  } -k   }
 { \mu+k(1 -  e^{ - \mu (x_0 +  \vert  x_0 \vert )  } )   } 
 \right]}   
    &~~\mathrm{if~~} x > 0  
  \end{array}
\right.
 \label{driftstarsurvivalinftyfinitedelta}
\end{eqnarray}
 while the conditioned killing rate of Eq. \ref{killingratestarsurvivalinftyfinite} reads
\begin{eqnarray}
k_{\infty}^*( x)  =  \frac{ \left( \frac{1- S^*(\infty) }{1- S( \infty \vert x_0)} \right)    }
{ Q_{\infty}^{[  S^*(\infty) ] }(0) } k \delta( x)
 =\frac{ [1- S^*(\infty) ] (k+\mu)  }
{   [1-S^*(\infty)]  
 + S^*(\infty) 
  \left[ \frac{  \mu  e^{ - \mu (x_0 +  \vert  x_0 \vert )  }  }
 { \mu+k(1 -  e^{ - \mu (x_0 +  \vert  x_0 \vert )  } )   } 
 \right] }  \delta ( x)
\label{killingratestarsurvivalinftyfinitedeltamu}
\end{eqnarray}

Let us mention the two special cases :

(i) When the conditioning is towards full-survival $S^*(\infty)=1$,
the conditioned drift of Eq. \ref{driftstarsurvivalinftyfinitedelta}
reduces to
\begin{eqnarray}
 \mu_{\infty}^*( x)  = \mu \left[ 1 +
\frac{   k  (1+ {\rm sgn}(x) )    }
{  (\mu +k ) e^{  \mu (x +  \vert  x \vert )  } -  k   } \right]
= 
 \left\lbrace
  \begin{array}{lll}
    \mu  
    &~~\mathrm{if~~} x < 0
    \\
    \mu \left[ 1 +
\frac{ 2  k      }
{  (\mu +k )e^{ 2 \mu x } -  k    } \right]
    &~~\mathrm{if~~} x> 0  
  \end{array}
\right.
 \label{driftstarsurvivalinftyfinitedelta1}
\end{eqnarray}

\begin{figure}[h]
\centering
\includegraphics[width=4.8in,height=4.2in]{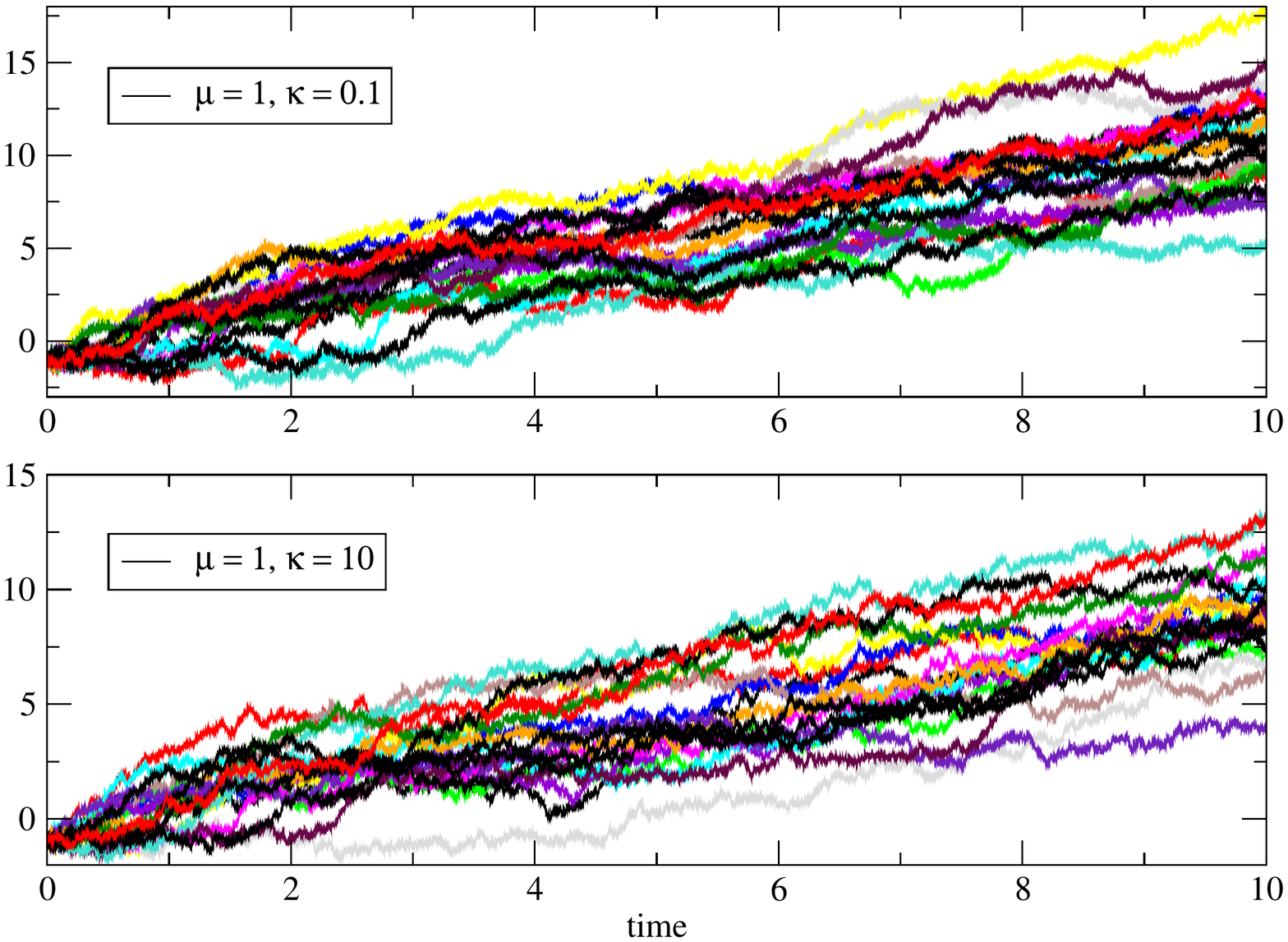}
\setlength{\abovecaptionskip}{15pt}  
\caption{A sample of $20$ diffusions satisfying the drift given by Eq. \ref{driftstarsurvivalinftyfinitedelta1} for two different values of the absorbing parameter $k$. Each color corresponds to the realization of one process. The starting point is $-1$ and the time step used in the discretization is $dt = 10^{-4}$. Simulations show that the conditioned process is not sensitive to the parameter $k$ (except around $x=0$). As $X^*(t)$ increases, the process behaves like a simple Brownian motion with drift $\mu$. }
\label{fig5}
\end{figure}

(ii) When the conditioning is towards zero-survival $S^*(\infty)=0$,
the conditioned drift of Eq. \ref{driftstarsurvivalinftyfinitedelta}
reduces to 
\begin{eqnarray}
 \mu_{\infty}^*( x)  =  - \mu \  {\rm sgn}(x) 
 = 
 \left\lbrace
  \begin{array}{lll}
    \mu  
    &~~\mathrm{if~~} x< 0
    \\
    - \mu
    &~~\mathrm{if~~} x>0  
  \end{array}
\right.
  \label{driftstarsurvivalinftyfinitedelta0}
\end{eqnarray}
while the conditioned killing rate of Eq. \ref{killingratestarsurvivalinftyfinitedeltamu} reads
\begin{eqnarray}
k_{\infty}^*( x)  =  (k+\mu)   \delta ( x)
\label{killingratestarsurvivalinftyfinitedelta0}
\end{eqnarray}

It is worth noticing that the stochastic process described by Eq. \ref{driftstarsurvivalinftyfinitedelta0}, without killing, is a Brownian motion with alternating drift or bang-bang process \cite{BorodinHandbook,touchette_bang_bang} and represents the dynamics of a Brownian particle in a symmetric wedge potential
\begin{eqnarray}
	V(x)  = \mu \vert  x \vert ~~~~~ \mu > 0
\label{wedgepotential}
\end{eqnarray}
Such a process has been introduced by de Gennes for studying dry friction \cite{deGennes}.


\section{Conclusion}

\label{sec_conclusion}

In this paper, we have revisited the conditioning of diffusion processes 
with space-dependent killing rates, in order to give a global discussion of the various conditioning constraints 
that can be imposed for finite horizon $T$ or for infinite horizon $T=+\infty$.
Firstly, we have characterized the conditioned process when one imposes both surviving distribution
$P^*(\vec y,T ) $ at time $T$
and the killing-distribution $K^*(\vec x_d,t_d ) $  for the intermediate times $t_d \in [0,T]$. Secondly, we have focused on  cases where the conditioning constraints are less-detailed than these full distributions $\left[ P^*(.,T) ; K^*(.,.)  \right] $, and we have constructed the appropriate conditioned processes via the optimization of the dynamical large deviations at Level 2.5 in the presence of the conditioning constraints that one wishes to impose. We have also analyzed the consequences for the limit of the infinite horizon $T \to +\infty$. Finally, we have described the application of this general construction to the pure diffusion in dimension $d$ with the quadratic killing rate $k(\vec x)= \gamma \vec x^2$, as well as to Brownian motion with uniform drift $\mu$ submitted to the delta killing rate $k(x)=k \delta(x)$ localized at the origin $x=0$, in order to generate stochastic trajectories satisfying various types of conditioning constraints.


\appendix

\section{Links with the dynamical large deviations at Level 2.5 and the stochastic control theory}

\label{app_largedev}

In this Appendix, one follows the Schr\"odinger perspective \cite{Schrodinger}
(see the recent detailed commentary \cite{CommentSchrodinger} accompanying its english translation
as well as in the two reviews \cite{ControlSchrodinger,MongeSchrodinger}).
The goal is then to analyze the large deviations properties
of a
 large number $N$ of independent realizations  $\vec X_n(t)$ of the unconditioned process
 labelled by $n=1,2,..,N$ starting all at the same initial condition $\vec X_n(0)=\vec x_0$.

\subsection{ Empirical ensemble-averaged observables for $N$ independent unconditioned processes $X_n(t)$ }

The ensemble-averaged density ${\hat P}(\vec x,t) $ at position $\vec x$ and at time $t$
 \begin{eqnarray}
 {\hat P}(\vec x,t) \equiv  \frac{1}{N} \sum_{n=1}^N \delta^d(\vec  X_n(t) - \vec x) 
\label{empiP}
\end{eqnarray}
follows some empirical dynamics that can be written as a continuity equation 
 \begin{eqnarray}
 \partial_t {\hat P}(\vec x,t) = -  \vec \nabla . { \vec {\hat  J}}(\vec x,t) - {\hat K}(\vec x,t)
\label{empidyn}
\end{eqnarray}
involving the empirical killing probability ${\hat K}(\vec x,t) $
and the empirical current ${ \vec {\hat  J}}(\vec x,t) $, that can be parametrized in terms of the empirical drift 
${ \vec {\hat  \mu }}(\vec x,t) $, while the diffusion coefficient $D (\vec x) $ is fixed
\begin{eqnarray}
{ \vec {\hat  J}}(\vec x,t) = { \vec {\hat  \mu }}(\vec x,t)  {\hat P}(\vec x,t) -   \vec \nabla  \left[ D (\vec x) {\hat P}(\vec x,t) \right]
\label{empiJ}
\end{eqnarray}
The normalization of the empirical density $ {\hat P}(\vec x,t) $ 
gives the empirical survival probability 
${\hat S}(t )$ at time $t$
\begin{eqnarray}
{\hat S}(t ) \equiv \int d^d \vec x {\hat P}(\vec x,t)   
\label{empiS}
\end{eqnarray}
whose time-decay involves the empirical killing probability ${\hat K}(\vec x,t) $
\begin{eqnarray}
 - \frac{  d {\hat S}(t ) }{dt}  = - \int d^d \vec x \partial_t {\hat P}(\vec x,t)   =   \int d^d \vec x  {\hat K}(\vec x,t)
 \label{empigammaK}
\end{eqnarray}

In the thermodynamic limit $N \to +\infty$, all these empirical observables concentrate 
on their typical values given by the corresponding observables without hats described 
in section \ref{sec_unconditioned} of the main text.
For large finite $N$, the dynamical fluctuations around these typical values 
can be analyzed via the large deviations at Level 2.5, as discussed
 in the next subsection.


\subsection{ Large deviations at Level 2.5 for the empirical dynamics during the time-window 
$t \in [0,T]$}

In the field of large deviation theory
(see the reviews \cite{oono,ellis,review_touchette} and references therein), 
the emergence of the Level 2.5 
describing the joint distribution of the empirical density and of the empirical flows
has been a major achievement \cite{chetrite_conditioned,chetrite_optimal,chetrite_HDR}. Indeed, in contrast to the Level 2 involving the empirical density alone,
the Level 2.5 can be written explicitly for general Markov processes,
including discrete-time Markov chains
 \cite{fortelle_thesis,fortelle_chain,review_touchette,c_largedevdisorder,c_reset,c_inference},
continuous-time Markov jump processes
\cite{fortelle_thesis,fortelle_jump,maes_canonical,maes_onandbeyond,wynants_thesis,chetrite_formal,BFG1,BFG2,chetrite_HDR,c_ring,c_interactions,c_open,c_detailed,barato_periodic,chetrite_periodic,c_reset,c_inference,c_runandtumble,c_jumpdiff,c_skew,c_metastable,c_east,c_exclusion}
and Diffusion processes 
\cite{wynants_thesis,maes_diffusion,chetrite_formal,engel,chetrite_HDR,c_reset,c_lyapunov,c_inference,c_metastable}.

In our present setting, 
the application of the large deviations at Level 2.5 
leads to the following conclusion :
the joint probability ${\cal P}^{[2.5]}_{[0,T]} \left[ {\hat P }(.,.) ; { \vec {\hat  J}}(.,.) ; {\hat K}(.,.) \right] $ to see the empirical density 
$ {\hat P}(\vec x,t) $, the empirical current ${ \vec {\hat  J}}(\vec x,t) $ and the empirical killing distribution ${\hat K}(\vec x,t) $
during the time window $0 \leq t \leq T$
follows the large deviation form for large $N$
\begin{eqnarray}
{\cal P}^{[2.5]}_{[0,T]} \left[ {\hat P }(.,.) ; { \vec {\hat  J}}(.,.) ; ; {\hat K}(.,.) \right]  \opsimeq_{N \to +\infty} 
{\cal C}^{[2.5]}_{[0,T]} \left[ {\hat P }(.,.) ; { \vec {\hat  J}}(.,.) ; {\hat K}(.,.) \right]
e^{- N {\cal I}^{[2.5]}_{[0,T]} \left[ {\hat P }(.,.) ; { \vec {\hat  J}}(.,.) ; {\hat K}(.,.) \right] }
\label{Level2.5}
\end{eqnarray}
with the following notations :

(i)  the rate function ${\cal I}^{[2.5]}_{[0,T]} \left[ {\hat P }(.,.) ; { \vec {\hat  J}}(.,.) ; {\hat K}(.,.) \right] $ at Level 2.5 
contains two contributions 
\begin{eqnarray}
{\cal I}^{[2.5]}_{[0,T]} \left[ {\hat P }(.,.) ; { \vec {\hat  J}}(.,.) ; {\hat K}(.,.)  \right]
&& = \int_0^T dt
\int_{-\infty}^{+\infty}  \frac{d x}{ 4 D (\vec x)  {\hat P}(\vec x,t) } \left[  { \vec {\hat  J}}(\vec x,t) -  \vec \mu  (\vec x) {\hat P}(\vec x,t) + \vec \nabla \left( D (\vec x) {\hat P}(\vec x,t) \right) \right]^2
\nonumber \\
&&    +\int d^d \vec x \int_0^T dt \left[  {\hat K}(\vec x,t)
      \ln \left(  \frac{{\hat K}(\vec x,t)  }{ k (\vec x) {\hat P}(\vec x,t) } \right) - {\hat K}(\vec x,t) +k (\vec x) {\hat P}(\vec x,t) \right]
\label{rate2.5}
\end{eqnarray} 
The first contribution involving the Ito drift $ \vec \mu (\vec x)$ and the diffusion coefficient $D (\vec x)$ 
corresponds to the usual rate function for diffusion processes \cite{wynants_thesis,maes_diffusion,chetrite_formal,engel,chetrite_HDR,c_reset,c_lyapunov,c_inference,c_metastable},
while the second contribution involving the killing rate $k (\vec x)$ 
corresponds to the usual rate function for Markov jump processes \cite{fortelle_thesis,fortelle_jump,maes_canonical,maes_onandbeyond,wynants_thesis,chetrite_formal,BFG1,BFG2,chetrite_HDR,c_ring,c_interactions,c_open,c_detailed,barato_periodic,chetrite_periodic,c_reset,c_inference,c_runandtumble,c_jumpdiff,c_skew,c_metastable,c_east,c_exclusion}:
 here the jump empirical flows associated to killing occur from position $\vec x$ towards the dead state.

(ii) the constitutive constraints ${\cal C}^{[2.5]}_{[0,T]} \left[ {\hat P }(.,.) ; { \vec {\hat  J}}(.,.) \right] $ 
at Level 2.5 can be decomposed into the initial condition $ {\hat P}(\vec x,t=0) = \delta^d(\vec x-\vec x_0)$
and the empirical dynamics of Eq. \ref{empidyn}
\begin{eqnarray}
{\cal C}^{[2.5]}_{[0,T]} \left[ {\hat P }(.,.) ; { \vec {\hat  J}}(.,.) ; {\hat K}(.,.) \right]
= \delta( {\hat P}(\vec x,0) - \delta^d(\vec x-\vec x_0) ) \ 
\prod_{t \in [0,T]} \prod_{\vec x} \left( \partial_t {\hat P}(\vec x,t) = -  \vec \nabla { \vec {\hat  J}}(\vec x,t) - {\hat K}(\vec x,t)
\right)
  \label{C2.5}
\end{eqnarray}

The parametrization of Eq. \ref{empiJ} allows to replace the empirical current ${ \vec {\hat  J}}(\vec x,t) $
by the empirical drift
\begin{eqnarray}
 { \vec {\hat  \mu }}(\vec x,t) = \frac{ { \vec {\hat  J}}(\vec x,t) +  \vec \nabla  \left[ D (\vec x) {\hat P}(\vec x,t) \right] }{ {\hat P}(\vec x,t) }
\label{empiJtomu}
\end{eqnarray}

As a consequence, the large deviations at Level 2.5 of Eq. \ref{Level2.5}
can be directly translated into the joint probability ${\cal P}^{[2.5]}_{[0,T]} \left[ {\hat P }(.,.) ; { \vec {\hat  \mu }}(.,.) ; {\hat K}(.,.) \right] $
to see the empirical density 
$ {\hat P}(\vec x,t) $, the empirical drift ${ \vec {\hat  \mu }}(\vec x,t) $ and the empirical killing distribution ${\hat K}(\vec x,t) $
\begin{eqnarray}
{\cal P}^{[2.5]}_{[0,T]} \left[ {\hat P }(.,.) ; { \vec {\hat  \mu }}(.,.) ; {\hat K}(.,.) \right] \opsimeq_{N \to +\infty} 
{\cal C}^{[2.5]}_{[0,T]} \left[ {\hat P }(.,.)  ; { \vec {\hat  \mu }}(.,.) ;  {\hat K}(.,.) \right]
e^{- N {\cal I}^{[2.5]}_{[0,T]} \left[ {\hat P }(.,.) ; { \vec {\hat  \mu }}(.,.) ; {\hat K}(.,.) \right] }
\label{Level2.5mu}
\end{eqnarray}
The rate function of Eq. \ref{rate2.5} translates into 
\begin{eqnarray}
{\cal I}^{[2.5]}_{[0,T]} \left[ {\hat P }(.,.) ; { \vec {\hat  \mu }}(.,.) ; {\hat K}(.,.) \right]
&& = \int_0^T dt
\int d^d \vec x {\hat P}(\vec x,t) \frac{ \left[  { \vec {\hat  \mu }}(\vec x,t)  -  \vec \mu  (\vec x)  \right]^2}{ 4 D (\vec x)   }
\nonumber \\
&&    +\int d^d \vec x \int_0^T dt \left[  {\hat K}(\vec x,t)
      \ln \left(  \frac{{\hat K}(\vec x,t)  }{ k (\vec x) {\hat P}(\vec x,t) } \right) - {\hat K}(\vec x,t) +k (\vec x) {\hat P}(\vec x,t) \right]
\label{rate2.5diffmu}
\end{eqnarray} 
while the constitutive constraints of Eq. \ref{C2.5} become
\begin{eqnarray}
&& {\cal C}^{[2.5]}_{[0,T]} \left[ {\hat P }(.,.) ; { \vec {\hat  \mu }}(.,.) ;  {\hat K}(.,.)  \right]
 = \delta( {\hat P}(\vec x,0) - \delta^d(\vec x-\vec x_0) ) 
\nonumber \\
&& \times
\prod_{t \in [0,T]} \prod_{\vec x} \left( \partial_t {\hat P}(\vec x,t) 
+  \vec \nabla \left[ { \vec {\hat  \mu }}(\vec x,t)  {\hat P}(\vec x,t) \right] -    \Delta  \left[ D (\vec x) {\hat P}(\vec x,t)  \right]+ {\hat K}(\vec x,t)  
\right)
  \label{C2.5diffmu}
\end{eqnarray}


\subsection{ Link with the stochastic control theory }

In this subsection, one assumes that the empirical density $ {\hat P}(\vec x,T) $ at time $T$ is given
and that the empirical killing probability ${\hat K}(\vec x,t) $ is given for $0 \leq t \leq T$
\begin{eqnarray}
  {\hat P}(\vec x,T) && = P^*(\vec x,T) 
  \nonumber \\
  {\hat K}(\vec x,t) && = K^*(\vec x,t)  \ \ \ \ \ \ \ {\rm for }  \ \ \ \ \ \ t \in [0,T]
  \label{empiT}
\end{eqnarray}


\subsubsection{ Optimization problem }

The goal is then to optimize the rate function ${\cal I}^{[2.5]}_{[0,T]} \left[ {\hat P }(.,.) ; { \vec {\hat  \mu }}(.,.); K^*(.,.)   \right] $ at Level 2.5 of Eq. \ref{rate2.5diffmu}
over the empirical density ${\hat P}(\vec x,t) $ and over the empirical drift ${ \vec {\hat  \mu }}(\vec x,t)  $
at all the intermediate times $t \in ]0,T[$, 
in the presence of the constitutive constraints of Eq. \ref{C2.5diffmu}
and the supplementary constraints of Eq. \ref{empiT}.
These constraints can be separated into :

(i) the boundary conditions for the empirical density $ {\hat P }(.,.) $
at the initial time $t=0$ and at the final time $t=T$ 
\begin{eqnarray}
 {\hat P}(\vec x,t=0) && = \delta^d(\vec x-\vec x_0) 
 \nonumber \\
   {\hat P}(\vec x,t=T) &&  = P^*(\vec x,T)  
 \label{timeboundaries}
\end{eqnarray}

(ii) the empirical dynamics for $t \in [0,T]$
\begin{eqnarray}
  \partial_t {\hat P}(\vec x,t) = -  \vec \nabla \left[ { \vec {\hat  \mu }}(\vec x,t)  {\hat P}(\vec x,t) \right] +    \Delta  \left[ D (\vec x) {\hat P}(\vec x,t) \right]
  - K^*(\vec x,t)
\label{bulkempi}
\end{eqnarray} 

As a consequence, for the intermediate times $ t \in ]0,T[$,
one only needs to optimize the rate function ${\cal I}^{[2.5]}_{[0,T]} \left[ {\hat P }(.,.) ; { \vec {\hat  \mu }}(.,.) ; K^*(.,.) \right] $ at Level 2.5 of Eq. \ref{rate2.5diffmu}
in the presence of the empirical dynamics (ii),
via the introduction of the Lagrangian
\begin{eqnarray}
{\cal L}\left[ {\hat P }(.,.) ; { \vec {\hat  \mu }}(.,.)  \right] 
 = {\cal I}^{[2.5]}_{[0,T]} \left[ {\hat P }(.,.) ; { \vec {\hat  \mu }}(.,.) ; K^*(.,.) \right]
+ {\cal L}^{Empi}\left[ {\hat P }(.,.) ; { \vec {\hat  \mu }}(.,.)  \right] 
\label{lagrangiantot}
\end{eqnarray} 
where the contribution 
\begin{eqnarray}
{\cal L}^{Empi}\left[ {\hat P }(.,.) ; { \vec {\hat  \mu }}(.,.)  \right] 
\equiv  \int_0^T dt
\int d^d \vec x \ \psi(\vec x,t) \left( \partial_t {\hat P}(\vec x,t) 
+  \vec \nabla . \left[ { \vec {\hat  \mu }}(\vec x,t)  {\hat P}(\vec x,t) \right] -    \Delta  \left[ D (\vec x) {\hat P}(\vec x,t) \right] + K^*(\vec x,t)
\right)
\ \ 
\label{lagrangianpsi}
\end{eqnarray} 
contains the Lagrange multiplier $\psi(\vec x,t)$ introduced in order to impose
the empirical dynamics of Eq. \ref{bulkempi}.


\subsubsection{ The adjoint-equation method }

As usual in stochastic control theory,
it is simpler to make some transformation of the Lagrangian before its optimization. 
In our present case, this amounts to rewrite the Lagrangian contributions of Eq. \ref{lagrangianpsi}
via integrations by parts,
 either over time $t \in ]0,T[$ using the boundary-conditions of Eq. \ref{timeboundaries}
\begin{eqnarray}
\int_0^T dt \ \psi(\vec x,t) \partial_t {\hat P}(\vec x,t) 
&& = \left[ \psi(\vec x,t)  {\hat P}(\vec x,t)\right]_{t=0}^{t=T} - \int_0^T dt {\hat P}(\vec x,t) \partial_t  \psi(\vec x,t)
\nonumber \\
&& =  \psi(\vec x,T)  P^*(\vec x,T) -  \psi(\vec x,0) \delta^d(\vec x-\vec x_0)- \int_0^T dt {\hat P}(\vec x,t) \partial_t  \psi(\vec x,t)
\label{integtime}
\end{eqnarray} 
or over space $\vec x$
\begin{eqnarray}
 \int d^d \vec x \ \psi(\vec x,t) 
\left(   \vec \nabla . \left( { \vec {\hat  \mu }}(\vec x,t)  {\hat P}(\vec x,t)  \right) -  \Delta  \left[ D (\vec x) {\hat P}(\vec x,t) \right] \right)
  =  - \int d^d \vec x {\hat P}(\vec x,t) \left[   { \vec {\hat  \mu }}(\vec x,t)    .  \vec \nabla\psi(\vec x,t) + D (\vec x)   \Delta\psi(\vec x,t)  \right]
  \ \ \ \ 
\label{integspace}
\end{eqnarray} 
Putting everything together, the Lagrangian of Eq. \ref{lagrangianpsi} reads
\begin{eqnarray}
  {\cal L}^{Empi}\left[ {\hat P }(.,.) ; { \vec {\hat  \mu }}(.,.)  \right] && =
 \int d^d \vec x \left[  \psi(\vec x,T)  P^*(\vec x,T) -  \psi(\vec x,0) \delta^d(\vec x-\vec x_0)- \int_0^T dt {\hat P}(\vec x,t) \partial_t  \psi(\vec x,t)\right]
\nonumber \\
&& - \int_0^T dt \int d^d \vec x {\hat P}(\vec x,t) \left[   { \vec {\hat  \mu }}(\vec x,t)   .   \vec \nabla\psi(\vec x,t) + D (\vec x)   \Delta\psi(\vec x,t)  \right]
 + \int_0^T dt  \int d^d \vec x \psi(\vec x,t) K^*(\vec x,t)
 \nonumber \\
 && =- \int_0^T dt  \int d^d \vec x {\hat P}(\vec x,t) \left[ \partial_t  \psi(\vec x,t) 
 + { \vec {\hat  \mu }}(\vec x,t)    .  \vec \nabla\psi(\vec x,t) + D (\vec x)   \Delta\psi(\vec x,t) 
 \right]
 \nonumber \\
 && + \int_0^T dt  \int d^d \vec x \psi(\vec x,t) K^*(\vec x,t)
 +  \int d^d \vec x  \psi(\vec x,t)  P^*(\vec x,t)  -  \psi(\vec x_0,0)   
\label{lagrangianpsiinteg}
\end{eqnarray} 
so that the lagrangian of Eq. \ref{lagrangiantot} becomes 
using the explicit rate function at Level 2.5 of Eq. \ref{rate2.5diffmu} 
\begin{eqnarray}
 {\cal L}\left[ {\hat P }(.,.) ; { \vec {\hat  \mu }}(.,.)  \right] 
&& = \int d^d \vec x  \psi(\vec x,T)  P^*(\vec x,T)  -  \psi(\vec x_0,0)   
\nonumber \\
 &&
+ \int_0^T dt \int d^d \vec x {\hat P}(\vec x,t) 
 \left( \frac{ \left[  { \vec {\hat  \mu }}(\vec x,t)  -  \vec \mu  (\vec x)  \right]^2}{ 4 D (\vec x)   }
  -  \left[ \partial_t  \psi(\vec x,t) 
 + { \vec {\hat  \mu }}(\vec x,t)   .   \vec \nabla\psi(\vec x,t) + D (\vec x)   \Delta\psi(\vec x,t) 
 \right] \right)
 \nonumber \\
&&    +\int d^d \vec x \int_0^T dt \left[  K^*(\vec x,t)
      \ln \left(  \frac{K^*(\vec x,t)  }{ k (\vec x) {\hat P}(\vec x,t) } \right) - K^*(\vec x,t) +k (\vec x) {\hat P}(\vec x,t) 
      +\psi(\vec x,t) K^*(\vec x,t) \right]
\label{lagrangianbulk}
\end{eqnarray} 
The optimization of Eq. \ref{lagrangianbulk}
over the empirical drift ${ \vec {\hat  \mu }}(\vec x,t) $
\begin{eqnarray}
0 && = \frac{ {\cal L}\left[ {\hat P }(.,.) ; { \vec {\hat  \mu }}(.,.)  \right] }{ \partial { \vec {\hat  \mu }}(\vec x,t) } = 
 {\hat P}(\vec x,t)  \left( \frac{   { \vec {\hat  \mu }}(\vec x,t)  -  \vec \mu  (\vec x)  }{ 2 D (\vec x)   }
 -    \vec \nabla\psi(\vec x,t) \right)
 \label{lagrangianbulkderimu}
\end{eqnarray} 
yields the optimal empirical drift ${ \vec {\hat  \mu }}^{opt}(\vec x,t) $ in terms of the Lagrange multiplier $\psi(\vec x,t)  $
\begin{eqnarray}
  { \vec {\hat  \mu }}^{opt}(\vec x,t)  =  \vec \mu  (\vec x) + 2 D (\vec x)    \vec \nabla\psi(\vec x,t)
 \label{mupsi}
\end{eqnarray} 
The further optimization of Eq. \ref{lagrangianbulk}
over the empirical density ${\hat P}(\vec x,t) $ reads using the optimal drift of Eq. \ref{mupsi}
\begin{eqnarray}
0 && = - \frac{ {\cal L}\left[ {\hat P }(.,.) ; { \vec {\hat  \mu }}(.,.)  \right] }{ \partial {\hat P }(\vec x,t) } 
\nonumber \\
&& = 
 - \frac{ \left[  { \vec {\hat  \mu }}^{opt}(\vec x,t)  -  \vec \mu  (\vec x)  \right]^2}{ 4 D (\vec x)   }
 +  \partial_t  \psi(\vec x,t) 
 + { \vec {\hat  \mu }}^{opt}(\vec x,t)    .  \vec \nabla\psi(\vec x,t) + D (\vec x)   \Delta\psi(\vec x,t) 
  + \frac{K^*(\vec x,t)  }{  {\hat P}(\vec x,t) } -k (\vec x)
 \nonumber \\
&& =  - D (\vec x)  \left[   \vec \nabla\psi(\vec x,t)  \right]^2
 +  \partial_t  \psi(\vec x,t) 
 + \left( \vec \mu  (\vec x) + 2 D (\vec x)    \vec \nabla\psi(\vec x,t) \right)  .   \vec \nabla\psi(\vec x,t) + D (\vec x)   \Delta\psi(\vec x,t) 
 + \frac{K^*(\vec x,t)  }{  {\hat P}(\vec x,t) } -k (\vec x)
     \nonumber \\
&& =   \partial_t  \psi(\vec x,t)  +  \vec \mu  (\vec x)    .    \vec \nabla\psi(\vec x,t) + D (\vec x)   \Delta\psi(\vec x,t) 
  + D (\vec x) \left[     \vec \nabla\psi(\vec x,t) \right]^2
  + \frac{K^*(\vec x,t)  }{  {\hat P}(\vec x,t) } -k (\vec x)
\label{lagrangianbulkderiP}
\end{eqnarray} 
The change of variables 
\begin{eqnarray}
 \psi(\vec x,t)  = \ln q(\vec x,t)
 \label{psiQ}
\end{eqnarray} 
transforms the non-linear Eq. \ref{lagrangianbulkderiP} for the Lagrange multiplier $\psi(\vec x,t)$
into the following linear backward dynamics for the function $q(\vec x,t)$
\begin{eqnarray}
- \partial_t  q(\vec x,t)    =    \vec \mu  (\vec x)  .  \vec \nabla  q(\vec x,t) + D (\vec x)    \Delta  q(\vec x,t) 
- k (\vec x) q(\vec x,t) + \frac{K^*(\vec x,t)  }{  {\hat P}^{opt}(\vec x,t) } q(\vec x,t)
\label{backwardsmallq}
\end{eqnarray} 
Using Eq. \ref{psiQ},
the optimal empirical drift ${ \vec {\hat  \mu }}^{opt}(\vec x,t) $ of Eq. \ref{mupsi} becomes
\begin{eqnarray}
  { \vec {\hat  \mu }}^{opt}(\vec x,t)  =  \vec \mu  (\vec x) + 2 D (\vec x)    \vec \nabla  \ln q(\vec x,t)
   \label{mupsiQ}
\end{eqnarray} 
while the optimal empirical density ${\hat P }^{opt}(\vec x,t) $ should be the solution of the corresponding empirical forward dynamics of Eq. \ref{bulkempi}
\begin{eqnarray}
 \partial_t {\hat P}^{opt}(\vec x,t) && =
  -   \vec \nabla . \left[ { \vec {\hat  \mu }}^{opt}(\vec x,t)  {\hat P}^{opt}(\vec x,t) \right] +    \Delta  \left[ D (\vec x) {\hat P}^{opt}(\vec x,t) \right] -K^*(\vec x,t)
  \nonumber \\
 && =
  -   \vec \nabla \left[ \left(  \vec \mu  (\vec x) + 2 D (\vec x)    \vec \nabla  \ln q(\vec x,t)\right) {\hat P}^{opt}(\vec x,t) \right] 
  +    \Delta  \left[ D (\vec x) {\hat P}^{opt}(\vec x,t) \right] -K^*(\vec x,t)
\label{forwardhat}
\end{eqnarray} 

Using the backward dynamics of Eq. \ref{backwardsmallq} for the function $q(\vec x,t)$
and the forward optimal dynamics of Eq. \ref{forwardhat} for ${\hat P}^{opt}(\vec x,t) $,
one obtains that the ratio
\begin{eqnarray}
  p(\vec x,t)  \equiv \frac{{\hat P}^{opt}(\vec x,t)}{ q(\vec x,t)}
   \label{defsmallp}
\end{eqnarray} 
satisfies the forward unconditioned dynamics involving the unconditioned operator ${\cal F}^{\dagger} $
of Eq. \ref{adjoint}
\begin{eqnarray}
  \partial_t p(\vec x,t)  && = \frac{ 1}{ q(\vec x,t)} \partial_t {\hat P}^{opt}(\vec x,t) - \frac{{\hat P}^{opt}(\vec x,t)}{ q^2(\vec x,t)} \partial_t q(\vec x,t)
  \nonumber \\
  && = -   \vec \nabla . \left[  \vec \mu (\vec x) p(\vec x,t)  \right] +   \Delta \left[ D (\vec x) p(\vec x,t)  \right] - k (\vec x) p(\vec x,t) 
  = {\cal F}^{\dagger}  p(\vec x,t)
   \label{forwardsmallp}
\end{eqnarray} 


\subsubsection{ Taking into account the time boundary conditions to obtain the final optimal solution }

In summary, the optimal solution ${\hat P}^{opt}(\vec x,t) $ is given the product of Eq. \ref{defsmallp}
\begin{eqnarray}
 {\hat P}^{opt}(\vec x,t) =  q(\vec x,t) p(\vec x,t)
   \label{optimalprod}
\end{eqnarray} 
where $p(\vec x,t)$ satisfies the forward unconditioned dynamics of Eq. \ref{forwardsmallp},
while $q(\vec x,t)$ satisfies the backward dynamics of Eq. \ref{backwardsmallq} that becomes using Eq. \ref{optimalprod}
\begin{eqnarray}
- \partial_t  q(\vec x,t)    =    \vec \mu  (\vec x)    \vec \nabla  q(\vec x,t) + D (\vec x)    \Delta  q(\vec x,t) 
- k (\vec x) q(\vec x,t) + \frac{K^*(\vec x,t)  }{  p(\vec x,t) } 
\label{backwardsmallqsmallp}
\end{eqnarray} 
In addition, we have to take into account
the time-boundary-conditions of Eq. \ref{timeboundaries}
 at the initial time $t=0$ and at the final time $t=T$
\begin{eqnarray}
 \delta^d(\vec x-\vec x_0)  && = {\hat P}^{opt}(\vec x,t=0) =  q(\vec x,0) p(\vec x,0)
 \nonumber \\
 P^*(\vec x,T) &&  = {\hat P}^{opt}(\vec x,t=T) =  q(\vec x,T) p(\vec x,T)
 \label{timeboundariessmall}
\end{eqnarray}

For the function $p(\vec x,t)$, it is natural to choose the unconditioned propagator $P(\vec x, t \vert \vec x_0,0)$
that would be the solution if we were not imposing atypical constraints
\begin{eqnarray}
p(\vec x,t) = P( \vec x,t \vert \vec x_0,0)
 \label{smallpBigP}
\end{eqnarray} 
With this choice, the backward dynamics of Eq. \ref{backwardsmallq}
becomes
\begin{eqnarray}
- \partial_t  q(\vec x,t)    =    \vec \mu  (\vec x)  .  \vec \nabla  q(\vec x,t) + D (\vec x)    \Delta  q(\vec x,t) 
- k (\vec x) q(\vec x,t) + \frac{K^*(\vec x,t)  }{   P( \vec x,t \vert \vec x_0,0) } 
\label{backwardsmallqbigP}
\end{eqnarray} 
while Eq. \ref{timeboundariessmall}
yields the following time boundary conditions for the function $q(\vec x,t)$ 
\begin{eqnarray}
 q(\vec x,t=0) && =1
 \nonumber \\
 q(\vec x,t=T) && = \frac{P^*(\vec x,t) }{ P( \vec x,t \vert \vec x_0,0)} 
 \label{timeboundariessmallq}
\end{eqnarray} 

The solution $q(\vec x,t)$
of the backward dynamics of Eq. \ref{backwardsmallq}
that satisfies the boundary conditions of Eqs \ref{timeboundariessmallq}
thus coincides with the function $Q(\vec x,t)$ introduced in Eq. \ref{Qdef} of the main text,
\begin{eqnarray}
 q(\vec x,t)  = Q(\vec x,t) 
 \label{smallqbigQ}
\end{eqnarray} 
and the optimal solution $ {\hat P}^{opt}(\vec x,t) $ of Eq. \ref{optimalprod}
 coincides with the conditioned probability $ P^*(\vec x,t)  $ introduced in Eq. \ref{conditional} in the main text
\begin{eqnarray}
 {\hat P}^{opt}(\vec x,t) =  q(\vec x,t) p(\vec x,t) =  Q(\vec x,t)P( \vec x,t \vert \vec x_0,0) = P^*(\vec x,t)
   \label{optimalprodstar}
\end{eqnarray}


\subsubsection{ Corresponding optimal value of the Lagrangian }

Using the second line of
Eq. \ref{lagrangianbulkderiP} to replace
\begin{eqnarray}
 \frac{ \left[  { \vec {\hat  \mu }}^{opt}(\vec x,t)  -  \vec \mu  (\vec x)  \right]^2}{ 4 D (\vec x)   }
-  \left[   \partial_t  \psi(\vec x,t)  + { \vec {\hat  \mu }}^{opt}(\vec x,t)      \vec \nabla\psi(\vec x,t) + D (\vec x)   \Delta\psi(\vec x,t) \right]
  =  \frac{K^*(\vec x,t)  }{  {\hat P}^{opt}(\vec x,t) } -k (\vec x)
\label{lagrangianbulkderiP2dline}
\end{eqnarray} 
one obtains the optimal value of the Lagrangian of Eq. \ref{lagrangianbulk} 
\begin{eqnarray}
 && {\cal L}\left[ {\hat P }^{opt}(.,.) ; { \vec {\hat  \mu }}^{opt}(.,.)  \right] 
   = \int d^d \vec x  \psi(\vec x,T)  P^*(\vec x,T)  -  \psi(\vec x_0,0)  
  \nonumber \\
 &&
+ \int_0^T dt \int d^d \vec x {\hat P}^{opt}(\vec x,t) 
 \left( \frac{ \left[  { \vec {\hat  \mu }}^{opt}(\vec x,t)  -  \vec \mu  (\vec x)  \right]^2}{ 4 D (\vec x)   }
  -  \left[ \partial_t  \psi(\vec x,t) 
 + { \vec {\hat  \mu }}^{opt}(\vec x,t)   .   \vec \nabla\psi(\vec x,t) + D (\vec x)   \Delta\psi(\vec x,t) 
 \right] \right)
 \nonumber \\
&&    +\int d^d \vec x \int_0^T dt \left[  K^*(\vec x,t)
      \ln \left(  \frac{K^*(\vec x,t)  }{ k (\vec x) {\hat P}^{opt}(\vec x,t) } \right) - K^*(\vec x,t) 
      +k (\vec x) {\hat P}^{opt}(\vec x,t) 
      + \psi(\vec x,t) K^*(\vec x,t)
      \right]
      \nonumber \\
&&   = \int d^d \vec x  \psi(\vec x,t)  P^*(\vec x,t)  -  \psi(\vec x_0,0)  
+ \int_0^T dt \int d^d \vec x {\hat P}^{opt}(\vec x,t) 
 \left( \frac{K^*(\vec x,t)  }{  {\hat P}^{opt}(\vec x,t) } -k (\vec x)  \right)
 \nonumber \\
&&    +\int d^d \vec x \int_0^T dt \left[  K^*(\vec x,t)
      \ln \left(  \frac{K^*(\vec x,t)  }{ k (\vec x) {\hat P}^{opt}(\vec x,t) } \right) - K^*(\vec x,t) +k (\vec x) {\hat P}^{opt}(\vec x,t) + \psi(\vec x,t) K^*(\vec x,t) \right]     
            \nonumber \\
&&   = \int d^d \vec x  \psi(\vec x,t)  P^*(\vec x,t)  -  \psi(\vec x_0,0)  
    +\int d^d \vec x \int_0^T dt  K^*(\vec x,t) \left[ \psi(\vec x,t) + 
      \ln \left(  \frac{K^*(\vec x,t)  }{ k (\vec x) {\hat P}^{opt}(\vec x,t) } \right)  \right]     
\label{lagrangianbulkopt}
\end{eqnarray} 
Using Eq. \ref{psiQ} and Eq. \ref{optimalprodstar}, the Lagrange multiplier 
\begin{eqnarray}
\psi(\vec x,t) = \ln  q(\vec x,t)= \ln Q(\vec x,t) = \ln \left( \frac{P^*(\vec x,t) }{P( \vec x,t \vert \vec x_0,0)} \right)
\label{psismallqbigQ}
\end{eqnarray} 
and its initial value
\begin{eqnarray}
\psi(\vec x_0,0) && = \ln q(\vec x_0,0) = \ln \left( \frac{P^*(\vec x_0,0) }{P(\vec x_0,0 \vert \vec x_0,0)} \right) = \ln ( 1 )=0
\label{lagrangianoptimalpsi}
\end{eqnarray} 
can be plugged into Eq. \ref{lagrangianbulkopt} to obtain that the optimal value of the Lagrangian
\begin{eqnarray}
 {\cal L}\left[ {\hat P }^{opt}(.,.) ; { \vec {\hat  \mu }}^{opt}(.,.)  \right] 
&& = 
  \int d^d \vec x    P^*(\vec x,T)  \ln \left( \frac{P^*(\vec x,T) }{P( \vec x,T \vert \vec x_0,0)} \right)
 \nonumber \\
&&        +\int d^d \vec x \int_0^T dt  K^*(\vec x,t) \left[ \ln \left( \frac{P^*(\vec x,t) }{P( \vec x,t \vert \vec x_0,0)} \right) + 
      \ln \left(  \frac{K^*(\vec x,t)  }{ k (\vec x) P^*(\vec x,t) } \right)  \right]   
      \nonumber \\
&& =  \int d^d \vec x    P^*(\vec x,T)  \ln \left( \frac{P^*(\vec x,T) }{P( \vec x,T \vert \vec x_0,0)} \right)
       +\int d^d \vec x \int_0^T dt  K^*(\vec x,t) 
      \ln \left(  \frac{K^*(\vec x,t)  }{ k (\vec x) P( \vec x,t \vert \vec x_0,0) } \right)       
      \ \ \  
\label{lagrangianbulkoptqfin}
\end{eqnarray} 
coincides with the Sanov rate function ${\cal I}^{Sanov}_T \left[ P^*(.,T) ; K^*(.,.)  \right] $
 given in Eq. \ref{RateSanovstar} of the main text, as it should for consistency.



\begin{thebibliography}{99}


\bibitem{refDoob} J.L. Doob, 
Bull. Soc. Math. Fr. 85, 431-48 (1957).

\bibitem{refbookDoob} J.L. Doob, {\em Classical Potential Theory and Its Probabilistic Counterpart}, Springer-Verlag, New York (1984).

\bibitem{refbookKarlin} S. Karlin and H. Taylor, 
{\em A Second Course in Stochastic Processes}, Academic Press, New York (1981).


\bibitem{refbookRogers} L.C.G. Rogers and D. Williams,  {\em Diffusions, Markov Processes and Martingales}, vol 2, Cambridge University Press, Cambridge (2000).


\bibitem{Borodin}
A.N. Borodin,  {\em Stochastic Processes}, Birkhauser, Springer International Publishing, Switzerland (2017).

\bibitem{refMajumdarOrland} S.N. Majumdar and H. Orland, 
J. Stat. Mech. P06039 (2015).


\bibitem{refHorne} J.S. Horne, E.O. Garton, S.M. Krone and J.S. Lewis, 
Ecology 88 (9), 2354-2363 (2007).

\bibitem{refBrody}  D.C. Brody, L.P. Hughston and A. Macrina,
 R. Elliott, M. Fu, R. Jarrow, J.Y. Yen (Eds.), Advances in Mathematical Finance, Festschrift vol. in honour of Dilip Madan, Springer (2007).
                  
\bibitem{refMulatier}  C. de Mulatier, E. Dumonteil, A. Rosso and A. Zoia, 
J. Stat. Mech. P08021 (2015). 

\bibitem{refbookPazsit} I. P\'azsit and L. P\'al, 
{\em Neutron Fluctuations: A Treatise on the Physics of Branching Processes}, Elsevier, Oxford (2008).


\bibitem{refMajumdarExcursion} S.N. Majumdar and A. Comtet, 
J. Stat. Phys. 119, 777-826 (2005).

\bibitem{refChung} K.L. Chung, 
Ark., Mat., 14, 155-177 (1976).

\bibitem{refMajumdarMeander} S.N. Majumdar , J. Randon-Furling,  M.J. Kearney and M. Yor, 
J. Phys. A, Math. Theor. 41, 365005 (2008).

\bibitem{refKnight} F.B. Knight, 
Trans. Amer. Soc. 73, 173–185 (1969).

\bibitem{refPinsky} R.G. Pinsky, 
Ann. Probab. 13 (2), 363-378 (1985).

\bibitem{refKorzeniowski} A. Korzeniowski, 
Stat. Probab. Lett. 8, 229 (1989).

\bibitem{refGarbaczewski} P. Garbaczewski, 
Phys. Rev. E 96 (3), 032104 (2017).

\bibitem{refAdorisio} M. Adorisio, A. Pezzotta, C. de Mulatier, C. Micheletti, and A. Celani, 
J. Stat. Phys. 170, 79-100  (2018).


\bibitem{refAlainTaboo} A. Mazzolo, 
J. Stat. Mech. P073204 (2018).

\bibitem{grela}
J. Grela, S.N. Majumdar and G. Schehr, J. Stat. Phys. 183, 1 (2021).

\bibitem{henri}
H. Orland, J. Chem. Phys. 134, 174114 (2011).



\bibitem{refSzavits} J. Szavits-Nossan and M.R. Evans,
J. Stat. Mech. P12008 (2015).

\bibitem{delarue}
M. Delarue, P. Koehl and H. Orland, J. Chem. Phys. 147, 152703 (2017).


\bibitem{refGarbaczewski_Levy} P. Garbaczewski and V. Stephanovich,
Phys. Rev. E 99, 042126 (2019).

\bibitem{bruyne_discrete}  
B. de Bruyne, S.N. Majumdar and G. Schehr, Phys. Rev. E 104, 024117 (2021).

\bibitem{Aguilar}
J. Aguilar, J.W. Baron, T. Galla and R. Toral, arXiv:2112.08252.

\bibitem{bruyne_run} 
B. de Bruyne, S.N. Majumdar and G. Schehr, J. Phys. A: Math. Theor. 54 385004 (2021).

\bibitem{refdeBruyne2022} 
B. de Bruyne, S.N. Majumdar and G. Schehr, Phys. Rev. Lett. 128, 200603 (2022)



\bibitem{refMazzoloJstat} A. Mazzolo, 
J. Stat. Mech. P023203 (2017).


\bibitem{Alain_OU}
 A. Mazzolo, J. Math. Phys. 58, 0953302 (2017).


\bibitem{refdeBruyne2021} B. de Bruyne, S.N. Majumdar, H. Orland and G. Schehr, 
J. Stat. Mech. 123204 (2021).

\bibitem{c_microcanonical}
C. Monthus, J. Stat. Mech. (2022) 023207.

\bibitem{us_LocalTime}
 A. Mazzolo and C. Monthus, arxiv:2205.15818.



\bibitem{Myers}
L.E. Myers, Journal of Applied Probability, 18,  523 (1981).

\bibitem{Berman}
S.M. Berman and H. Frydman, Communications in Statistics : Stochastic Models, 12:3, 367 (1996).

\bibitem{Holcman}
D. Holcman, A. Marchewka and Z. Schuss, math-ph/0502035.

\bibitem{Toste}
S. Toste and D. Holcman, arxiv: 2201.05915.


\bibitem{Karlin1982}
S. Karlin and S. Tavare, Stochastic Processes and their Applications 13, 249 (1982).
 
\bibitem{Karlin1983}
S. Karlin and S. Tavare, SIAM J. Appl. Math. 43, 31 (1983).


\bibitem{Frydman}
H. Frydman, Comm. Stat. Stochastic Models 16, 189 (2000).


\bibitem{Steinsaltz}
D. Steinsaltz and S.N. Evans, Trans. Amer. Math. Soc. 359, 1285 (2007).

\bibitem{kolb}
M. Kolb and D. Steinsaltz,  Annals of Probability 40, 162 (2012).

\bibitem{Evans2019}
S.N. Evans and A. Hening, Stoch. Process. Their Appl. 129, 1622 (2019).






\bibitem{tryphon_killing}
Y. Chen, T.T. Georgiou and M. Pavon,
arxiv:2108.02879.




\bibitem{Schrodinger}
E. Schr\"odinger, Sitzungsberichte der preussischen Akademie der Wissenschaften, 
physikalisch-mathematische Klasse, 8 N9, 144 (1931).

\bibitem{CommentSchrodinger}
R. Ch\'etrite, P. Muratore-Ginanneschi and K. Schwieger,
Eur. Phy. J. H 46, 28 (2021).

\bibitem{ControlSchrodinger}
Y. Chen, T.T. Georgiou and M. Pavon,
Journal of Optimization Theory and Applications 169, 671 (2016).

\bibitem{MongeSchrodinger}
Y. Chen, T.T. Georgiou and M. Pavon, SIAM 63, 249 (2021).




\bibitem{refBaudoin} F. Baudoin, 
Stoch. Proc. Appl. 100, 109-145 (2002). 


\bibitem{refMultiEnds} 
C. Larmier, A. Mazzolo and A. Zoia, 
J. Stat. Mech. (2019) 113208.

\bibitem{us_DoobFirstPassage} 
C. Monthus and A. Mazzolo, arxiv:2202.12047.

\bibitem{us_DoobFirstEncounter} 
A. Mazzolo and C. Monthus, J. Phys. A: Math. Theor. https://doi.org/10.1088/1751-8121/ac7af3 (2022).


\bibitem{gardiner}
C. W. Gardiner, 
{\em Handbook of Stochastic methods}, Springer-Verlag Berlin (1990).



\bibitem{oono}
Y. Oono,
Progress of Theoretical Physics Supplement 99, 165 (1989).

\bibitem{ellis}
R.S. Ellis, Physica D 133, 106 (1999).

\bibitem{review_touchette}
H. Touchette, Phys. Rep. 478, 1 (2009).


\bibitem{BorodinHandbook}
A.N. Borodin and P. Salminen, 
{\em Handbook of Brownian motion-facts and formulae}, Springer Science \& Business Media (2015).

\bibitem{touchette_bang_bang}
H. Touchette, E. V. der Straeten and W. Just, J. Phys. A: Math. Theor. 43 445002 (2010).

\bibitem{deGennes}
P.G. de Gennes, J. Stat. Phys. 119, 953 (2005).





\bibitem{chetrite_conditioned}
R. Ch\'etrite and H. Touchette
 Ann. Henri Poincare 16, 2005 (2015).

\bibitem{chetrite_optimal}
R. Ch\'etrite and H. Touchette, J. Stat. Mech. P12001 (2015).

\bibitem{chetrite_HDR}
R. Ch\'etrite, HDR Thesis (2018)
"P\'er\'egrinations sur les ph\'enom\`enes al\'eatoires dans la nature",
 Laboratoire J.A. Dieudonn\'e, Universit\' e de Nice.



\bibitem{fortelle_thesis}
A. de La Fortelle, PhD (2000)
"Contributions to the theory of large deviations and applications" INRIA Rocquencourt.

\bibitem{fortelle_chain}
G. Fayolle and A. de La Fortelle,
Problems of Information Transmission 38, 354 (2002).


\bibitem{c_largedevdisorder}
C. Monthus, Eur. Phys. J. B 92, 149 (2019).

\bibitem{c_reset}
C. Monthus, J. Stat. Mech. (2021) 033201.


\bibitem{c_inference}
C. Monthus, J. Stat. Mech. (2021) 063211. 



\bibitem{fortelle_jump}
A. de La Fortelle, 
Problems of Information Transmission 37, 120 (2001).



\bibitem{maes_canonical}
C. Maes and K. Netocny, Europhys. Lett. 82, 30003 (2008).

\bibitem{maes_onandbeyond}
C. Maes, K. Netocny and B. Wynants, Markov Proc. Rel. Fields. 14, 445 (2008).

\bibitem{wynants_thesis}
B. Wynants, arXiv:1011.4210, PhD Thesis (2010), "Structures of Nonequilibrium Fluctuations", Catholic University of Leuven.


\bibitem{chetrite_formal}
A.C. Barato and R. Ch\'etrite, J. Stat. Phys. 160, 1154 (2015).

\bibitem{BFG1}
L. Bertini, A. Faggionato and D. Gabrielli, 
Ann. Inst. Henri Poincare Prob. and Stat. 51, 867 (2015).

\bibitem{BFG2}
L. Bertini, A. Faggionato and D. Gabrielli, 
Stoch. Process. Appli. 125, 2786 (2015).


\bibitem{c_ring}
C. Monthus, J. Stat. Mech. (2019) 023206.

\bibitem{c_interactions}
C. Monthus, J. Phys. A: Math. Theor. 52, 135003 (2019).


\bibitem{c_open}
C. Monthus, J. Phys. A: Math. Theor. 52, 025001 (2019).

\bibitem{c_detailed}
C. Monthus, J. Phys. A: Math. Theor. 52, 485001 (2019).

\bibitem{barato_periodic}
A.C. Barato and R. Ch\'etrite, J. Stat. Mech. (2018) 053207.

\bibitem{chetrite_periodic}
L. Chabane, R. Ch\'etrite and G. Verley, J. Stat. Mech. (2020) 033208.

\bibitem{chabane_thesis}
L. Chabane, PhD Thesis (2021) "From rarity to typicality : the improbable journey of a large deviation",
Universit\'e Paris-Saclay.


 \bibitem{c_runandtumble}
C. Monthus, J. Stat. Mech. (2021) 083212.

\bibitem{c_jumpdiff}
C. Monthus,  J. Stat. Mech. (2021) 083205.

\bibitem{c_skew}
C. Monthus, J. Stat. Mech. (2021) 103202.

\bibitem{c_metastable}
C. Monthus, J. Stat. Mech. (2022) 013206.

\bibitem{c_east}
C. Monthus,  Eur. Phys. J. B 95, 32 (2022).

\bibitem{c_exclusion}
C. Monthus, J. Stat. Mech. (2021) 123205.




\bibitem{maes_diffusion}
C. Maes, K. Netocny and B.  Wynants
Physica A 387, 2675 (2008).


\bibitem{engel}
J. Hoppenau, D. Nickelsen and A. Engel,
 New J. Phys. 18 083010 (2016).

\bibitem{c_lyapunov}
C. Monthus, J. Stat. Mech. (2021) 033303.





\end{thebibliography}
\end{document}